\documentclass[twocolumn]{aastex631}

\usepackage[colorinlistoftodos]{todonotes}

\usepackage{rotating}
\usepackage{float}
\usepackage{amssymb}
\usepackage{comment}
\usepackage{multirow,acronym}
\usepackage{natbib}
\usepackage{listings} 
\usepackage{makecell}
\usepackage{booktabs}
\usepackage{amsmath}
\usepackage{subfigure}
\usepackage{color}
\usepackage{xcolor}
\usepackage{hyperref}
\usepackage[T1]{fontenc}
\usepackage{graphicx}
\usepackage{bm}
\usepackage{CJK} 
\hypersetup{colorlinks=true, citecolor=blue, 
  linkcolor=cyan, urlcolor=magenta}
\usepackage[linesnumbered,ruled]{algorithm2e}
\newcommand{\code}[1]{\lstinline{#1}}
\newcommand{\kratos}{\code{Kratos}}
\usepackage{aas_macros}
\usepackage{amsfonts}
\usepackage{float}
\usepackage{amsmath,amssymb}
\usepackage{natbib}    
\usepackage{hyperref} 
\usepackage{graphicx} 
\usepackage{color}
\usepackage{verbatim}
\usepackage{enumitem}
\usepackage{natbib}

\usepackage[linesnumbered,ruled]{algorithm2e}

\newcommand{\proptosim}{\mathrel{\vcenter{
 \offinterlineskip\halign{\hfil$##$\cr
 \propto\cr\noalign{\kern2pt}\sim\cr\noalign{\kern-2pt}}}}}
\renewcommand{\b}[1]{\boldsymbol{#1}}
\newcommand{\unit}[1]{{\rm #1}}

\newcommand{\response}[1]{#1}

\renewcommand{\min}{\mathrm{min}}
\renewcommand{\max}{\mathrm{max}}

\newcommand{\m}{\unit{m}}
\newcommand{\g}{\unit{g}}
\newcommand{\G}{\unit{G}}
\newcommand{\K}{\unit{K}}

\newcommand{\s}{\mathrm{s}}

\newcommand{\B}{\mathbf{B}}     
\newcommand{\E}{\mathbf{E}}     
\newcommand{\J}{\mathbf{J}}     
\renewcommand{\v}{\mathbf{v}}   

\renewcommand{\d}{\mathrm{d}}

\newcommand{\e}{\mathrm{e}}

\newcommand{\p}{\mathrm{p}}     

\renewcommand{\O}{\mathrm{O}}     
\newcommand{\A}{\mathrm{A}}     
\renewcommand{\H}{\mathrm{H}}     
\renewcommand{\P}{\mathrm{P}}     

\newcommand{\figdir}{.}

\begin{document}

\title{The \kratos{} Framework for Heterogeneous
  Astrophysical Simulations:\\ Fundamental Infrastructures
  and Hydrodynamics}

\author[0000-0002-6540-7042]{Lile Wang}
\affil{The Kavli Institute for Astronomy and Astrophysics,
  Peking University, Beijing 100871, China}
\affil{Department of Astronomy, School of Physics, Peking
  University, Beijing 100871, China}

\correspondingauthor{Lile Wang}
\email{lilew@pku.edu.cn}

\begin{abstract}
  The field of astrophysics has long sought computational 
  tools capable of harnessing the power of modern GPUs to
  simulate the complex dynamics of astrophysical
  phenomena. The \kratos{} Framework, a novel GPU-based
  simulation system designed to leverage heterogeneous
  computing architectures, is introduced to address these
  challenges.  \kratos{} offers a flexible and efficient
  platform for a wide range of astrophysical simulations, by
  including its device abstraction layer, multiprocessing
  communication model, and mesh management system that
  serves as the foundation for the physical module
  container. Focusing on the hydrodynamics module as an
  example and foundation for more complex simulations,
  optimizations and adaptations have been implemented for
  heterogeneous devices that allows for accurate and fast
  computations, especially the mixed precision method that
  maximize its efficiency on consumer-level GPUs while
  holding the conservation laws to machine accuracy. The
  performance and accuracy of \kratos{} are verified through a
  series of standard hydrodynamic benchmarks, demonstrating
  its potential as a powerful tool for astrophysical
  research.
\end{abstract}


\section{Introduction}
\label{sec:intro}

Astrophysical simulations are necessary to study a vast
array of phenomena, from the intricate dynamics of planetary
systems to the large-scale structure of the universe
itself. These simulations are inherently complex, requiring
the resolution of multiple physical processes across an
extraordinary range of spatial and temporal scales. To meet
these challenges, computational astrophysics has long relied
on advanced numerical methods and high-performance computing
resources.  The advent of graphical processing units (GPUs)
has revolutionized the field of computational sciences, by
offering the potential for substantial speedups in the
computational performance. However, the actual design of
GPUs, or heterogeneous computing devices as a kind, is {\it
  not} to simply speed up the computation task; instead,
these devices actually speeds up the overall computing
througput via adequately parallerizing the computing tasks.

In this context, \kratos{} is introduced as a novel
GPU-optimized simulation system designed to address the
challenges of heterogeneous simulations in astrophysics and
beyond. \kratos{} should represent the significant advancement
in the field that has been emerging since recent years till
the near future, offering a unified framework that supports
a wide range of physical modules, including hydrodynamics,
consistent thermochemistry, magnetohydrodynamics,
self-gravity, radiative processes, and more.  A critical
aspect in the development of \kratos{} is its performance
optimization for heterogeneous devices while maximizing the
flexibility. Because of the architechtures of GPUs are
special from the hardware to the programming model, detailed
designs are necessary for the infrastructures on which those
modules are implemented.  Its design philosophy centers on
achieving high performance without compromising accuracy,
ensuring that it remains a versatile tool for a diverse
array of astrophysical applications.

\kratos{} is among the handful of astrophysical fluid
dynamics codes that attempt to leverage the power of GPUs,
but it is neither the first nor the sole one. Prior works,
such as \citet{Grete_k_athena} for K-Athena and
\citet{2024arXiv240916053S} for AthenaK, have detailed the
integration of MHD solvers within Athena++ using Kokkos
\citep{kokkos}. \citet{2023A&A...677A...9L} have introduced
IDEFIX as a Kokkos-based adaptation of the PLUTO
code. Several other astrophysical fluid dynamics codes also
utilize GPU-accelerated computing through platform-specific
tools like CUDA, including Ramses-GPU
\citep{2017ascl.soft10013K}, Cholla
\citep{2015ApJS..217...24S}, and GAMER-2
\citep{2018MNRAS.481.4815S}. \response{ Nevertheless,
  \kratos{} continues to pursue outstanding performance by
  directly applying optimizations to low-level features that
  are closer to heterogeneous hardware, while maintaining
  flexibility through the use of modern computational
  programming practices. It is designed as a general-purpose
  framework for multiple physical modules, with the
  hydrodynamic modules detailed in this paper as the
  starting point.}

\kratos{} is composed in modern \code{C++}.  Instead of
building on existing high-performance heterogeneous
computing libraries that operate on higher levels (such as
Kokkos), \kratos{} complies with the \code{C++}-17 standard
and the corresponding standard library, and construct the
framework directly on the GPU programming model, e.g., CUDA
for NVIDIA GPUs \citep{cuda}, HIP for AMD GPUs and the
derivatives \citep{hip}, by composing an abstraction layer
\response{with zero runtime overhead} to insulate the actual
implementation of algorithms from the subtle details and
nuances existing in different programming models. \response{
  This approach guarantees the compatibility with different
  kinds of heterogeneous and conventional devices made by
  most mainstream manufacturers.  With the help of proper
  utilities (e.g.,
  HIP-CPU\footnote{\url{https://github.com/ROCm/HIP-CPU}}),
  \kratos{} also maintains the compatibility with the
  ``conventional'' CPU architechtures for developing
  purposes, and can be extended to actually operate on these
  architechtures when necessary.  Meanwhile, the
  higher-level encapsulations (including Kokkos) typically
  focus on the most generic compatibility and portability,
  which could cause significant performance impact due to
  the inability to fully include the low-level features that
  are not available on all supported platforms. In contrast,
  the algorithms directly implementated on low-level
  features in \kratos{} are necessary to optimize the code
  performance by maximizing the utilization of various
  heterogeneous features, especially detailed dispatching of
  computing streams and threads, as well as the shared
  memory allocated on the caches of multiprocessors (see
  e.g., \S\ref{sec:hydro}). This approach also enables deep
  utilization of template metaprogramming to maintain a
  high-level modularity and maximize the flexibility and
  capability of \kratos{} as a platform, on which more
  physical modules are to be included (detailed in incoming
  papers).
}  \kratos{} itself is based on the basic programming
models (e.g., CUDA or HIP) and the standard MPI (Message
Passing Interface) implementations.  Otherwise, it is fully
self-contained, rather than depending on third-party
libraries (including the input and output modules, and the
subsequent data reading and processing programs in Python),
in order to maximize the compatibility and ease the
deploying procedures on different platforms.

Before this paper describing the methods, \kratos{} has
already been used (after thorough tests) in several
astrophysical studies \citep[e.g.][]{2022ApJ...932..108W,
  2024ApJ...977..274L}, and multiple further applications
have also been carried out in various astrophysical
scenarios. \response{Multiple fundamental modules, including
  the input parameter parser, binary inputs and outputs
  module, and the device abstraction layer, have also been
  used and made publicly available in other codes composed
  or contributed by the author
  \citep[e.g.][]{2025ApJS..276...40W}.  } This paper,
describing the foundamental infrastructures for all modules
implemented on \kratos{} for various astrophysical systems,
also serves as the foundation of several forthcoming papers
on the method and modules, including MHD, thermochemistry
and particle-based simulations (L. Wang, in prep.).

The structure of this paper is as follows. A detailed
description of the basic infrastructure in
\S\ref{sec:infrastructure} is followed by the elaboration on
the implementation of the hydrodynamics module in
\S\ref{sec:hydro} as an example of general modules. A
comprehensive set of code tests, including accuracy,
symmetry, and performance optimizations, are included in
\S\ref{sec:tests}. Finally, \S\ref{sec:summary} summarizes
the paper, and outlines the incoming works for multiple
physical mechanisms based on the \kratos{} framework.

\section{Basic infrastructures}
\label{sec:infrastructure}

\begin{figure*}
  \centering
  \includegraphics[width=6.5in, keepaspectratio]
  {\figdir/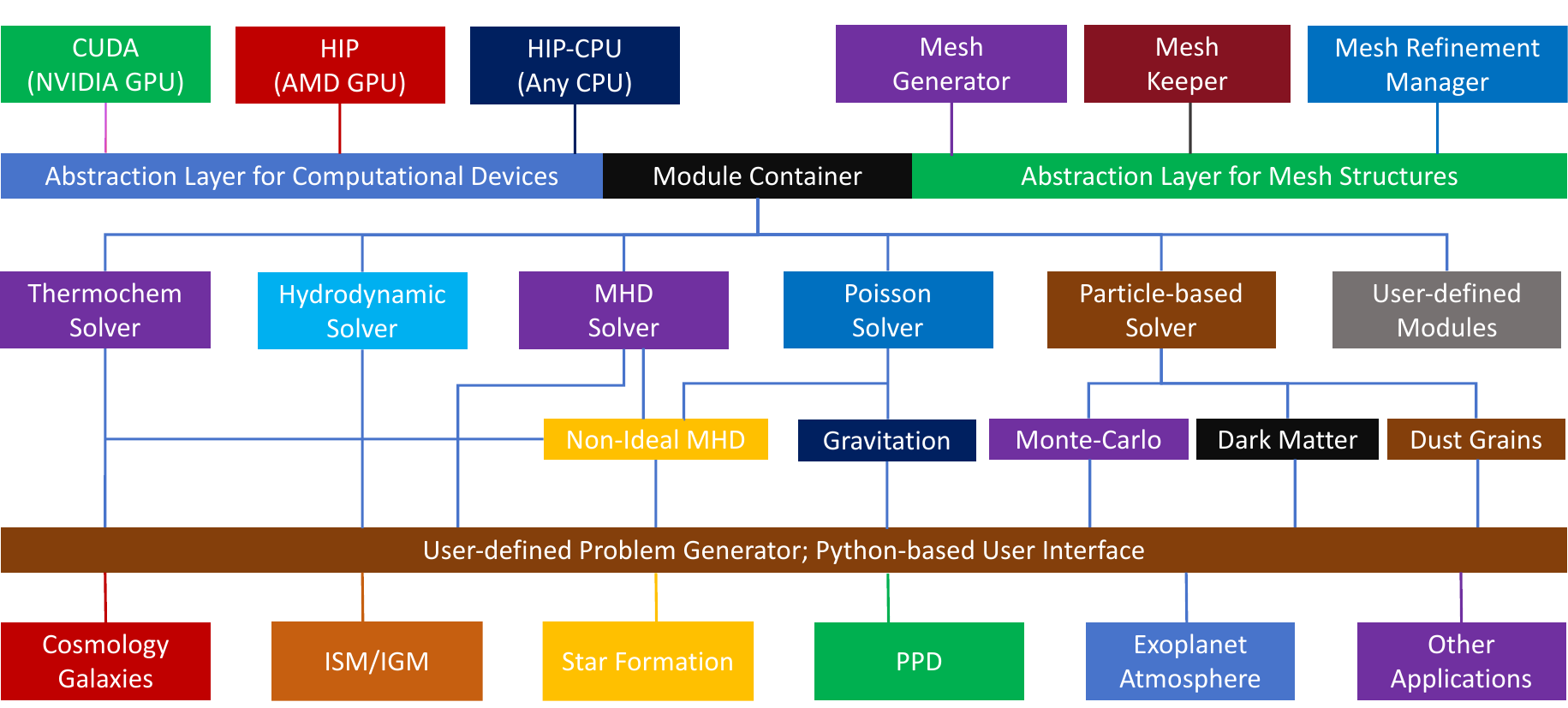}
  \caption{The overall structure of the \kratos{} code. Based
    on the abstration layer for computing devices and the
    abstraction layer for the mesh structures, the module
    container holds all capable modules for physical
    processes taking place in astrophysical simulations
    (hydrodynamic module elaborated in this paper; other
    modules will be described in forthcoming papers,
    L. Wang, in prep.). While complicated physical
    calculations can be constructed through the interactions
    of modules, a user interface (problem generator) handles
    the definition of intial, boundary, and in-simulation
    conditions. Such interface can also handle the initial
    conditions defined by a Python-based interface. Actual
    astrophysical objects and processes are simulated based
    on all fundamental systems described above.  }
  \label{fig:struct} 
\end{figure*}

The fundamental infrastructures of the \kratos{} code are
briefed in Figure~\ref{fig:struct}. One of the most apparent
feature it holds is the separation of the mesh structures
from the physical modules. In general, the mesh structure
sub-system only contains the geometry of the mesh, including
the overall mesh tree on which the sub-grids (``blocks''
hereafter) are allocated. The physical modules are included
in a separate module container, which has similar
functionality as the ``task lists'' in many other numerical
simulation systems \citep[e.g.][]{stone_athena_2020}, and
each module has an internal ``tast list'' of its own. The
execution sequence can be designed by users in each cycle of
\kratos{} runs for better flexibility and compatibility.

\subsection{Abstraction layer for devices}
\label{sec:infra-device-layer}

In what follows, the paper use the term ``host'' to denote
the hardware and operations on the CPU (including the RAM on
the CPU side), and ``device'' for the hardwares, software
interfaces, and instructions that are hetrogeneous to the
CPUs (e.g., GPUs; the term ``GPU'' will be used to denote
all hetrogeneous devices with similar architechtures in what
follows). On the contemporary high-performance and
consumer-level computing markets, there have already been
multiple programming models available, which are usually
specific to the manufacturers of devices. To extend the
portability on various hardware and software platforms while
maximizing the performance by keeping the code close to the
hardware, \kratos{} decides to implement the computations and
algorithms via a new pathway by formulating {\it all}
device-side codes in the subset that is the intersection
between HIP and CUDA syntaxes, avoiding the
usage of any manufacturer-specfic device calls.

On the host side that dispatches data transfer and device
function (namely ``kernels'') launching, the interfaces
could have subtle differences (e.g., \code{cudaMallocHost}
versus \code{hipHostMalloc} for allocating non-pagable host
memory spaces). To resolve similar issues, \kratos{} also
holds an abstraction layer on both host and device
sides. This abstraction layer is implemented as a \code{C++}
class, containing the following ingredients that encapsulate
the differences in syntax across various programming models:
\begin{itemize} 
\item Initialization and finalization (destruction and
  resource release) for device enviroments;
\item Device dispatching that assign a specific device to a
  CPU process;
\item Fundamental attributes, including \code{__global__},
  \code{__host__}, \code{__device__} etc., which are
  commonly used in almost all relevant programming models,
  including CUDA and HIP;
\item Directives related to shared memory on the device
  side;
\item Application interfaces, including memory space
  allocations and de-allocations, host-device data transfer
  and copy procedures, calls of device calls (namely
  ``device kenel launching''), and GPU stream and event
  operations.
\end{itemize}
All nuances comparing different programming models are
absorbed by the abstraction layer, which assures \kratos{}
to be compiled and run on different platforms without any
modifications to the code. \response{ Meanwhile, features
  that are particularly crucial for GPUs, such as shared
  memory allocation and access, constant memory access, as
  well as stream and event operations, remain readily
  accessible. } It is noticed that the intermediate layers
have been introduced to models other than HIP and CUDA,
including oneAPI \citep{oneapi} and OpenCL 3.0
\citep{opencl3}. Equipped with encapsulation models such as
chipStar
\footnote{\url{https://github.com/CHIP-SPV/chipStar}}, codes
based on \code{HIP} can also be compiled and run on systems
that are compatible with \code{oneAPI} and \code{OpenCL
  3.0}. What is more, using the open source library
\code{HIP-CPU} that emulates GPUs via the TBB (thread
building block) library, \kratos{} simulations can also be
conducted using CPU with various architechtures that are
compatible with TBB, including x86, ARM, and RISC-V.

\subsection{Multiprocessing communication model}
\label{sec:infra-comm}

\kratos{} can be operated on multiple devices in parallel, and
the code design requires that each device is controlled and
managed by one host-side process. Data transfer among
devices and processes are handeled by the communication
module, which is established on the MPI (message passing
interface) by default, while other implementations are also
possible. It is worth noting that several maufacturers have
introduced ``device-aware MPI'', which allows direct
transfers of device-side data without passing through the
host-side memory. In general, however, \kratos{} does {\it
  not} always utilize this ``device-awareness'' MPI feature,
because of the absence of ``streams'' in the
MPI interfaces disables asynchronous communication
\citep[e.g.][]{mpigpu}. \response{
  In typical simulations involving multiple blocks (or
  sub-meshes; see\S \ref{sec:infra-mesh-tree}),
  communication operations can be launched immediately once
  each block is ready, without the need to wait the
  completion of processing all blocks on a single GPU. The
  asynchronous pattern is thus essential for all possible
  efforts to effectively ``hide'' communication operations
  behind the computations of other blocks. Consequently,
  \kratos{} has chosen to develop its own independent
  communication module on which the mesh management system
  is constructed, instead of relying on existing AMR
  frameworks such as AMReX \citep{2019JOSS....4.1370Z}.
}

\kratos{} employs a unified and encapsulated multiprocessing
communication module, which adopts the standard MPI
implementation as its default ``backend,'' adhering to
version 3 of the standard. The module interfaces are largely
similar to the MPI API in the C programming language, with
an extra feature to optimize device-to-device
communications: interfaces such as non-blocking send and
receive operations are designed to accept device streams as
arguments. This allows a communication task to be queued on
a specific device stream, ensuring the correct handling of
data dependencies and executaion sequence. These operations
inevitably involves data transfer between host and device
memories, which causes extra time consumption of data
transfer. The introduction of device stream, however,
enables the procedure pattern that allows the data transfer
to take place following the correct sequence after all
prerequisite computations are acompished.  By properly
design the pattern of communication procedures, \kratos{}
generally attempts to maximize the efforts of ``hiding'' the
communication operations behind the computations, and thus
improving the overall performance of multi-device
computation. Note that such communication feature is {\it
  unlikely} to be implemented even with the device-aware MPI
(such as the ``CUDA-aware'' or ``HIP-aware'' MPI
implementations that allow direct communications between
device-side memory spaces), due to the absence of device
streams in device-aware MPI API parameters. Details of the
design for asynchronous communications utilizing streams are
described in the explanation of each model, including the
standard hydrodynamic module elaborated in
\S\ref{sec:hydro}.

The special communication module requires host-side buffers
for proper data transfer, which are designed to be allocated
on the first invocation. Capacities of these buffers are
given by multiplying the requested size by a safety factor,
which is typically around $\sim 1.2$. This factor has been
found to be effective in the majority of multiple scenarios,
although it should be noted that the optimal value may
differ depending on the specific application
requirements. These buffers are intentionally retained
rather than being released back to the system, unless the
subsequent requests for buffer sizes exceed their current
capacity (when the original buffer is released and replaced
with a newly allocated one), or until intentional
de-allocation (e.g., one simulation as a whole comes to an
end). Note that, although the hydrodynamic solver described
in this paper does not involve variable buffer sizes, this
feature is still required by other modules (especially
particle modules) elaborated in following papers.  This
design choice minimizes the time-consuming overhead of
buffer allocation and deallocation, thereby maintaining a
clean and efficient runtime environment.

\subsection{Mesh manager}
\label{sec:infra-mesh-manager}

\subsubsection{Mesh tree and refinement}
\label{sec:infra-mesh-tree}

\kratos{} utilizes $2^d$-trees to manage its structured mesh,
where $d \in \{ 1, 2, 3\}$ (thus octree in three dimensions,
quadtree in two dimensions, and binary tree in one
dimension). The trees fulfill dual roles: (1) to identify
the spatial configuration of the sub-grids, and (2) to
facilitate the allocation of geometry necessary for mesh
refinement. Each node within this tree (not to be confused
with the server node that denotes a single set of computer)
referred to as a ``block''. These blocks are the
foundational elements that collectively compose the entire
mesh, covering the entire domain of the simulation.

Based on the tree structure, mesh refinement is implemented
for \kratos{}. Each block is represented by a node in the
tree. Once a block is to be refined, it is replaced by $2^d$
blocks that have doubled resolution and halved size. When
the number of refinement level $L$ is greater than one, the
$2^d$-tree refinement scheme is carried out recursively,
with an extra procedure adding prospectively extra
refinement regions to guarantee that the absolute value of
the difference in $L$ is no more than one for neighboring
blocks.

The tree structure is organized by the \code{C++} standard
template library (STL) member \code{std::map} with the
four-integer key values allocated in lexicographical
order. These four-integers, indicated as $\{I, J, K, L\}$,
define the ``block indices'': the first three ($I, J, K$)
stand for the logical indices of the block, and the $L$ for
the refinement level (the coarsest level has $L = 0$).  When
launched in parallel, each process holds a copy of the basic
tree that holds the key values and communication information
(including the identity number and the process rank number)
of all blocks, while the detailed geometry data are only
filled into the nodes that are held by the current
process. This approach guarantees that all neighbors of
every block can be easilily identified for communication
purposes, while the tree size and communication costs are
minimized when the tree is updated.

\begin{figure*}
  \centering
  \includegraphics[width=6.5in, keepaspectratio]
  {\figdir/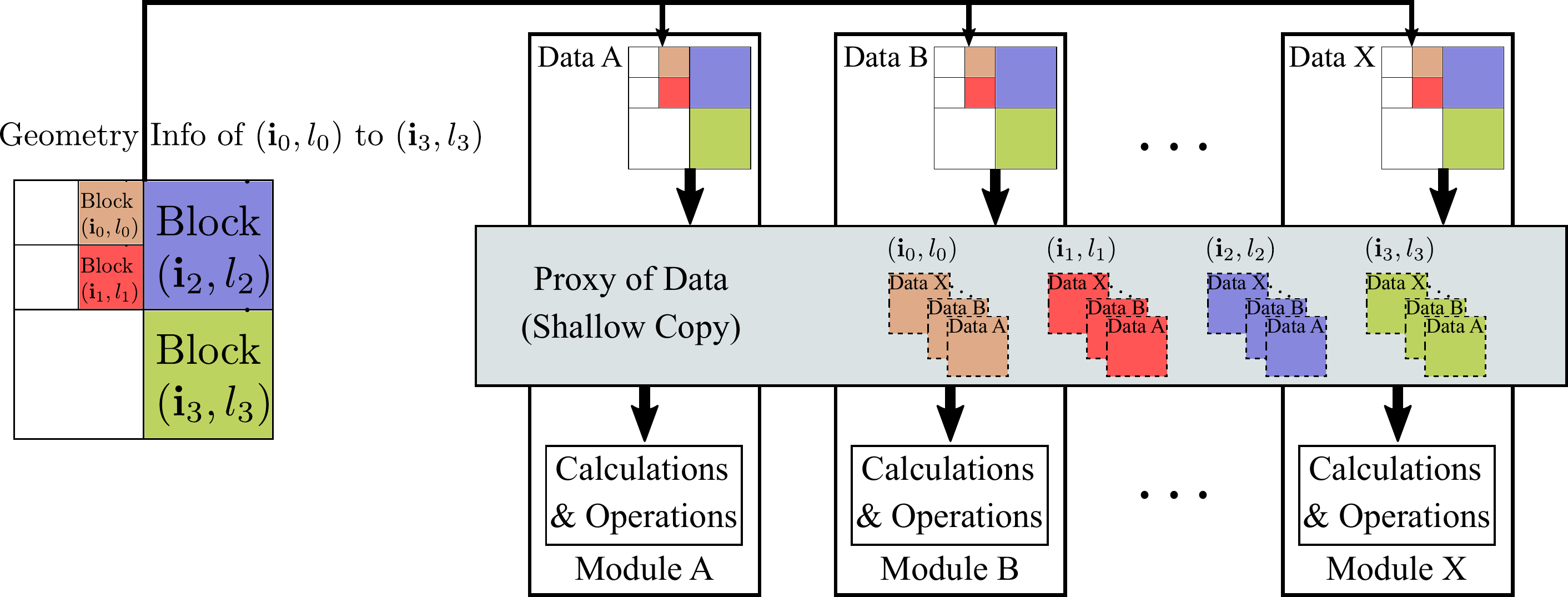}
  \caption{The method that \kratos{} handles the interaction
    of different physical processes via sharing the data
    held by modules. An instant data structure,
    encapsulating the shallow copy of all data needed held
    by all releva modules, is created as a data proxy right
    before the launching of each GPU job conducting the
    calculations. }
  \label{fig:proxy} 
\end{figure*}

\subsubsection{Job dispatching and load balancing}
\label{sec:infra-load-balance}

When \kratos{} simulations are carried out on multiple
devices, the computation load are dispatched to every device
involved based on the mesh tree stucture. The job
dispatching is managed by the load balancing module, which
requires a scheme that maps the blocks distributed in the
$d$-dimensional space onto a one-dimensional line
segment. Using this mapping scheme, blocks are eventually
distributed on the computing devices using the device
number, either evenly or by a user-defined array of weights
that takes other factors into account (e.g., the actual
computing capability of each device).  Although \kratos{}
allows users to enroll their own implementation, by default,
there are two different load balancing module that can be
selected from.

The first is the row-filling scheme, which maps the
$(I, J, K)$ block indices onto one dimension
lexicographically. When the $2^d$-tree mesh refinement is
applied, the simple row-filling scheme is applied to the
blocks in the refined region, and replace the ``original''
coarse block in-place. The row-filling scheme addresses
simulations in simple geometric layout with efficiency very
close to optimal schemes, and suits several situations even
better than other schemes when the row-filling scheme
minimizes cross-process data dependencies (e.g., in
simulations with plane-parallel ray tracing).

The second method is based on fractal space-filling curves,
which naturally resolves the ordering of the refined region,
since these curves can be generated recursively to a finer
level once refinement is applied. Among the possible
selections, the Hilbert curves ususally performs better for
several reasons. The Hilbert curves are 2-based, in contrast
to some other famous choices (e.g., the 3-based Peano
curves), which maximizes the flexibility of the system. In
addition, the Hilbert curve guarantees that the neighbors on
the mapped one-dimensional space are also neighbors in the
$d$-dimension space, which excels over other possibilities
e.g. Z-ordering curves.  To maximize the flexibility,
\kratos{} also allows users to adopt their own rules for
generating space-filling curves under the Lindenmayer System
(L-system for short), and the default rules for Hilbert
curves are also implemented under the L-system for 2D and
3D, respectively. In 3D, the L-system for Hilbert curves
reads,
\begin{equation}
  \label{eq:hilbert}
  \begin{split}
    & X \rightarrow
    \wedge<XF\wedge<XFX-F\wedge>> \\
    & \qquad \qquad XFX\&F+>>XFX-F>X-> \ ; \\
    & X \Rightarrow \text{Axiom (recursion interface)}\ ,\\
    & F \Rightarrow \text{Move forward}\ ,\\
    & + \Rightarrow \text{Yaw angle }+\pi/2\ ,\\
    & - \Rightarrow \text{Yaw angle }-\pi/2\ ,\\
    & \wedge \Rightarrow \text{Pitch angle }+\pi/2\ ,\\
    & \& \Rightarrow \text{Pitch angle }-\pi/2\ ,\\
    & > \Rightarrow \text{Roll angle }+\pi/2\ ,\\
    & < \Rightarrow \text{Roll angle }-\pi/2\ .
  \end{split}
\end{equation}
The application of space-filling curve load balancing will
presented when illustrating the examples of hydrodynamics
(\S\ref{sec:tests-outflow-amb}).


\subsection{Modules and module manager}
\label{sec:infra-module-manager}


Physical processes within a simulation are computed by
specialized physical modules managed by the module
container, which in turn relies on the data provided by the
mesh manager. Unlike most other simulation systems, the
relationship between the module container and the mesh
manager is unidirectional in \kratos{}: the mesh does not hold
any data or information about the modules, while the
physical modules extract the necessary geometric data from
the mesh to perform their calculations.  When simulations
are distributed across multiple devices, inter-block
communication often becomes necessary for different physical
modules to function effectively. Since each module may have
unique communication requirements and patterns, the modules
handle all related data and communication procedures
independently.  The data needed for these computations is
stored within the modules themselves. This approach
maximizes the flexibility of \kratos{}, which allows it to
work as a code defined {\it fully} by the modules
involved--for example, although hydrodynamics is the
foundation of many modules, \kratos{} can still be a radiative
transfer code that is totally irrelevant to the dynamics of
fluids (H., Yang and L. Wang, in prep.).

In scenarios where data must be shared among multiple
modules, explicit specification of data coupling is
employed, facilitated through a data proxy scheme. This
scheme involves creating a ``shallow copy'' of the data,
which includes only the essential geometries and pointers to
the GPU memory, on the CPU side before the module is
executed on the GPU side. This method is depicted in
Figure~\ref{fig:proxy}, and the implementation details of
the modules will be discussed within the context of the
modules themselves,providing a clearer understanding of how
they operate and interact within the system.

The modules included with Kratos are intentionally crafted
to be user-accessible for modification and expansion. The
typical approach to extending their functionality is through
inheritance, where users create custom classes by inheriting
from the base default \code{C++} class and overriding the
relevant member functions. However, it is important {\it
  not} to implement this overriding using runtime
polymorphism via virtual functions in
\code{C++}. \response{According to the tests conducted by
  the author in parallel to the developing procedures, the
  usage of virtual tables can often lead to significant
  performance degradation (by $\sim 10^2$ times) on most
  heterogeneous devices}\footnote{Typically this performance
  impact is attributed to the virtual tables allocated on
  the global graphics memory on GPUs, which could suffer
  from severe latency when applied.}, with various tests
indicating a decrease in performance by more than one order
of magnitude (not detailed in the current paper).  Moreover,
there is generally no practical need for runtime
polymorphism in simulations, as the methods and algorithms
used are typically predetermined.  Consequently, \kratos{}
opts for a different strategy, the compilation-time
polymorphism through template metaprogramming. This approach
utilizes the Curious Recursive Template Pattern (CRTP; see
also \citealt{crtp}), which is based on F-bounded
polymorphism \citep[see][] {F_bounded}. By employing this
pattern, \kratos{} can achieve the desired flexibility and
customization without incurring the performance penalties
associated with runtime polymorphism, thus ensuring that the
modules remain efficient and effective for simulation tasks.

\section{Hydrodynamics: the example and foundation of many
  modules}
\label{sec:hydro}

\begin{sidewaysfigure*}  
  \centering
  \vspace{10cm}    
  \includegraphics[width=9.4in, keepaspectratio]
  {\figdir/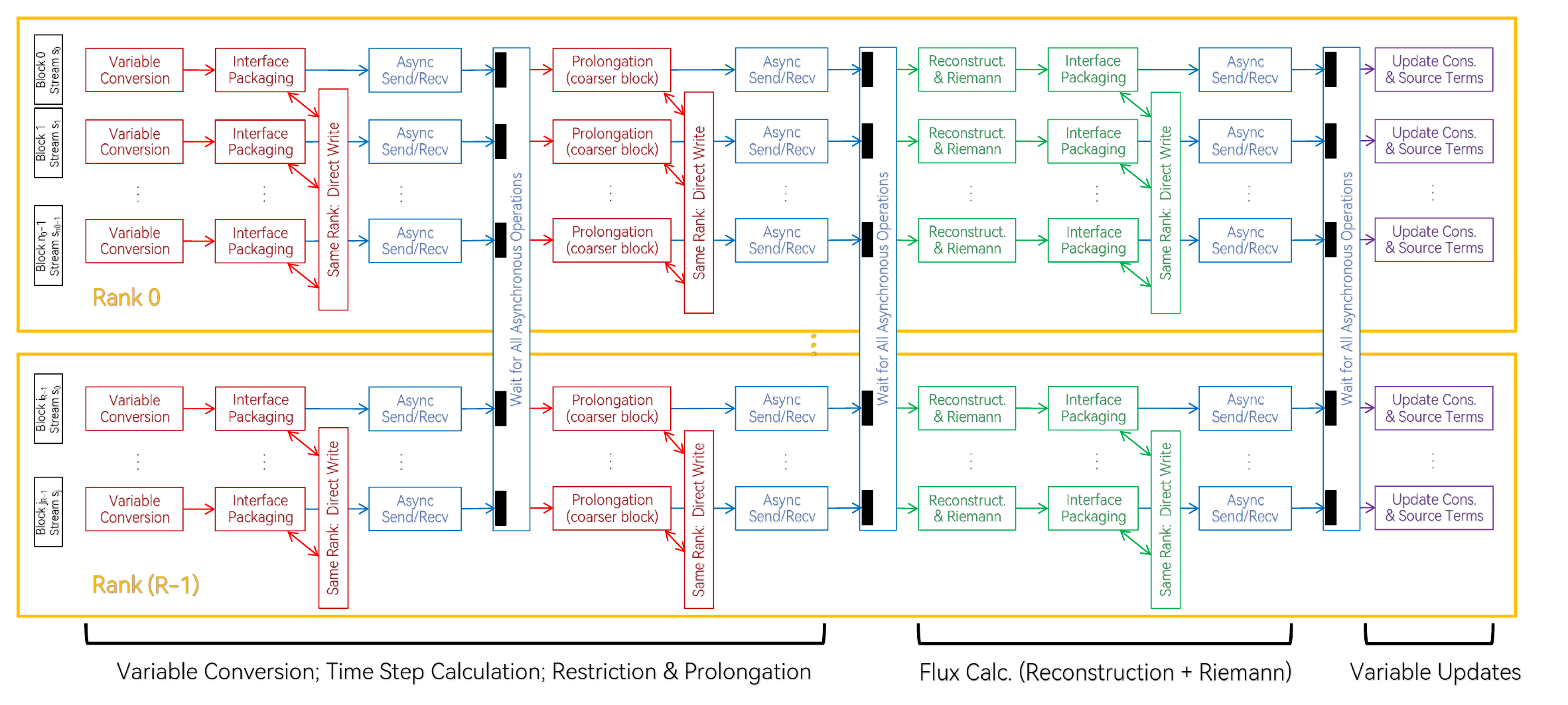}
  \caption{Scheme for one hydrodynamic cycle in \kratos{},
    showing the parallelization pattern and the interface
    calls for data exchange. Each ``rank'' denotes a
    process, typically controlling one GPU, on which the
    calculations for multiple blocks are conducted. Each row
    (labelled by the ``Block $X$, Stream $s_X$'' at the
    beginning of each row) denote the numerical processes of
    one block on the $2^d$-tree mesh, which are located on
    one GPU stream for proper data synchronizations.
    Different steps of the Godunov solver are marked by
    colors (red for the preparation of primative variabs,
    green for calculation the Riemann fluxes, and magenta
    for the updates of conservative variables), while the
    communication operations are marked with blue. Black
    solid blocks stand for the synchronizations of each
    stream, taking place within the calls of communication
    operations to minimize waiting and thus maximize the
    efficiency.}
  \label{fig:hydro_proc}
\end{sidewaysfigure*}

The default hydrodynamics module in \kratos{} is not only the
first implemented example module, but also the most
optimized one. It plays a pivotal role in \kratos{}, as
hydrodynamics is a fundamental aspect of various
astrophysical research disciplines. Moreover, it serves as
the cornerstone for numerous other modules and applications,
such as magnetohydrodynamics, reacting flows, radiative
hydrodynamics, and cosmology, among others. While detailed
discussions of these modules and applications will be
presented in subsequent papers, the focus of this work is to
delineate and clarify the adaptations and optimizations made
for heterogeneous devices. These optimizations are also the
methodological foundation for the forthcoming studies.

The default hydrodynamic module is constructed within the
Godunov framework, solving the Euler equation on descrete
mesh,
\begin{equation}
  \label{eq:hydro}
  \begin{split}
    & \partial_t \rho + \nabla \cdot (\rho \v) = 0\ ;
    \\
    & \partial_t (\rho \v) + \nabla \cdot \left(\rho \v \v +
      p \mathbf{I} \right) = 0 \ ;
    \\
    & \partial_t \epsilon + \nabla \cdot \left[
      \left(\epsilon + p
      \right) \v \right] = 0\ ,  
  \end{split}
\end{equation}
where $\rho$, $p$ and $\mathbf{v}$ are the fluid mass
density, pressure and velocity,
$\epsilon = \epsilon_{\rm g} + \rho v^2/2$ is the total
energy density ($\epsilon_{\rm g}$ is the gas internal
energy density), and $\partial_t$ denotes time derivatives.
The solution procedures can be broken down into three key
``sub-modules'': (1) the slope-limited Piecewise Linear
Method (PLM) reconstruction, (2) the Harten-Lax van Leer
with contact surface (HLLC) Riemann solver, and (3) the Heun
time integrator, which is one of the second-order
Runge-Kutta methods. When mesh refinement is employed, it is
crucial to implement specialized schemes at the boundaries
between finer and coarser mesh levels. The algorithms that
govern the fluxes across cell surfaces are mathematically
equivalent to those found in most existing computational
fluid dynamics codes, as referenced in
\citep{toro_riemann_2009}. However, to achieve optimized
performance, it is essential to tailor these algorithms
specifically for heterogeneous devices via re-arranging the
sub-modules and schemes summarized in
Figure~\ref{fig:hydro_proc}. The synchronization calls of
streams, which mark and guarantee that all operations are
finished in sequence prior to the call on the same stream,
are ``postponed'' to the communication. Because there are
{\it no} dependency among different streams until
synchronizations, such pattern allows the actual data
transfer (from the GPU to the CPU side, and from one process
to another) to take place in the background while the GPU is
handling the prerequisite calculations for other blocks. In
this way, a relatively large fraction of the communication
costs are hidden behind the computation, so that the
communication time is minimized. Detailed elaborations of
more optimizations will follow in the subsequent discussion.




\subsection{Variable conversion and boundary conditions}
\label{sec:hydro-cons2prim}

The default Riemann solvers in \kratos{} hydrodynamics work
with primative variables--for hydrodynamics, the variable
set contains $(\rho, p, \mathbf{v})$. As the conservative
variables are the basic ones whose evolution is calculated
on each step, conversion from conservative to primative
variables needs to be calcualted first.  The length of
timestep $\Delta t$ is also calculated in the meantime of
such variable conversion based on the
Courant-Lewvy-Freidrich (CFL) conditions, which is conducted
only on the first substep when the time integrator adopts a
multi-step method. 

If a Godunov algorithm has $n$th-order spatial accuracy, the
reconstruction step will require $n$ cells on both sides for
each cell interface. The consistency of calculations require
primative variables to be available in the ghost zones,
which form ``buffer layers'' surrounding blocks. This paper
refers to the schemes of setting physical variables in the
ghost zones as ``boundary conditions'', which include both
``physical boundaries'' (p-boundaries) for the actual
boundary of the whole simulation domain, and ``communication
boundaries'' (c-boundaries) for the interfaces between
neighbouring blocks.

\subsubsection{Physical boundary conditions}
\label{sec:hydro-p-bnd}

\begin{figure}
  \centering 
  \includegraphics[width=2.8in, keepaspectratio]
  {\figdir/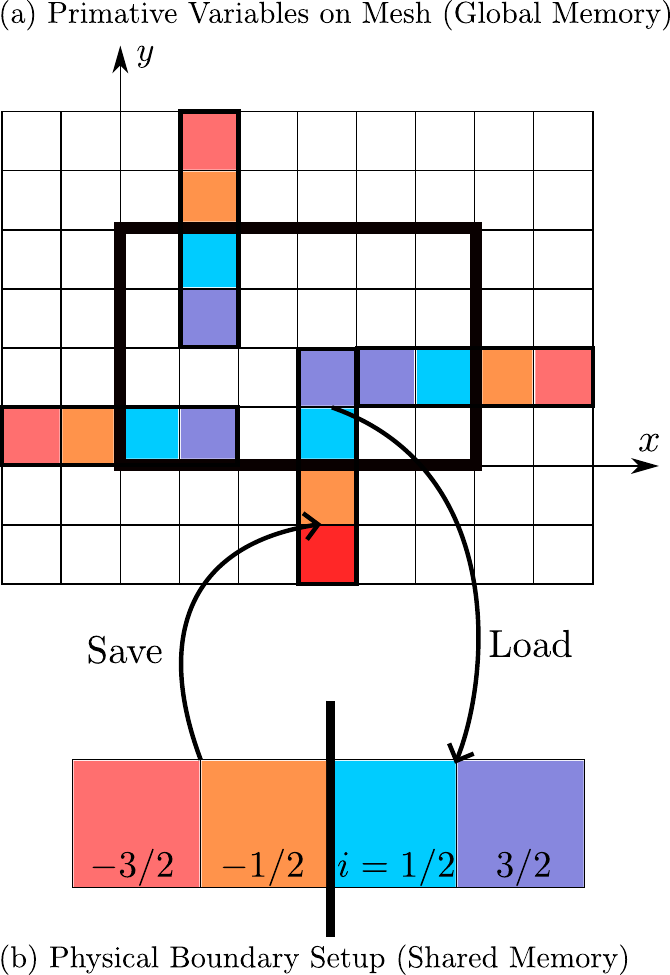}
  \caption{Boundary condition methods for hydrodynamics,
    showing four normal columns as examples to illustrate
    how \kratos{} rotate the normal columns to the $x$
    direction to unify the methods on the interfaces of
    block boundaries.  Colors indicate the corresponding
    cells before and after the rotations.  }
  \label{fig:bnd_conv}
\end{figure}

\kratos{} allows users to enroll their own p-boundary
conditions to accommodate for the actual situations of
astrophysical simulations. Note that when periodic boundary
conditions are applied to the domain, the domain boundaries
become c-boundaries instead. On the programming side, the
p-boundary module in \kratos{} conducts proper manimupations
including the reflection and alternation of coordinate axes
during the loading and saving stages of the variables inside
and near the p-boundaries, so that all actual p-boundary
operations are mathematically equivalent to dealing with the
boundary condition in the $x$-direction on the left, as is
shown in Figure~\ref{fig:bnd_conv}.

Users can set different p-boundaries on different sides.
There are three default options, which are expressed with
the indexing for the left
$x$-boundary (see also Figure~\ref{fig:bnd_conv}; 
integers are used for cell faces, and half integers for cell
centers):
\begin{enumerate}
\item Free boudnary: $q_{-i-1/2} = q_{1/2}$;
\item Outflow boundary: $q_{-i-1/2} = q_{1/2}$,\\
  but $v_{x, -i-1/2} = \min\{v_{x, 1/2}, 0\}$;
\item Reflecting boudnary: $q_{-i-1/2} = q_{i+1/2}$, \\
  but $v_{x, -i-1/2} = -v_{x,i+1/2}$.
\end{enumerate}
Here $0\leq i \leq n_{\rm gh} - 1$, and $n_{\rm gh}$
corresponds to the spatial order of the reconstruction. Note
that the transform automatically rotates the normal
component of velocity (pointing inwards) into $v_{x}$, and
rotates $v_x$ back to the desired component and direction
after the boundary condition algorithms are applied.

\subsubsection{Communication boundaries}
\label{sec:hydro-c-bnd}

\begin{figure*}
  \centering
  \hspace*{-0.4cm}
  \includegraphics[width=6.5in, keepaspectratio]
  {\figdir/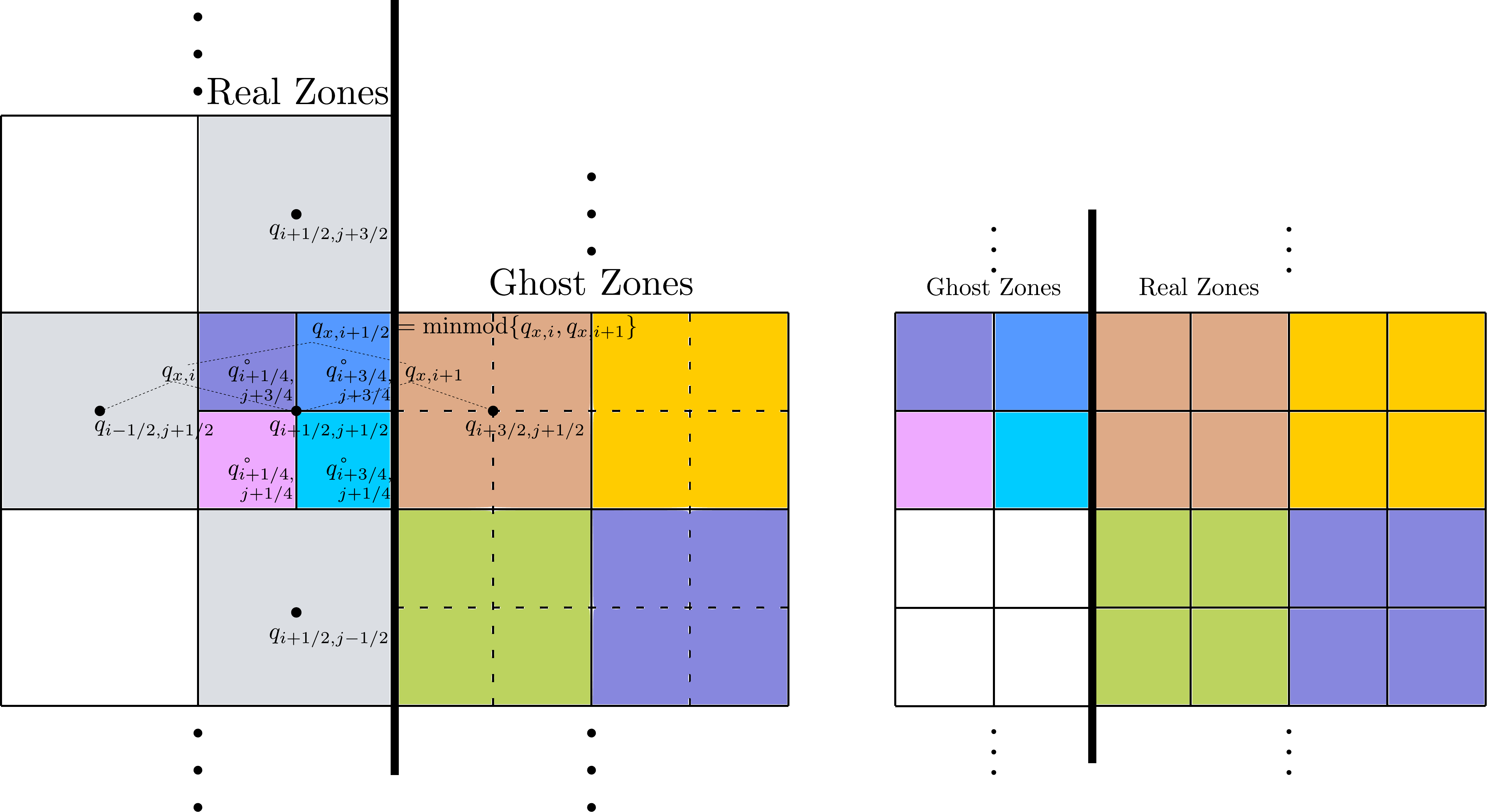}
  \caption{Restriction and prolongation operations near the
    block refinement interface. Colors are used to mark the
    corresponding relations across the interface separating
    coarser (left) and finer (right) blocks. The
    prolongations are computed on the coarser side, and the
    $x-$ derivatives $q_{x,I,J}$ are calculated using minmod
    slope limiter in a pattern indicated on the corser mesh
    (derivatives along other directions can be obtained
    similarly).  }
  \label{fig:hydro_refine} 
\end{figure*}


When a simulation domain is partitioned into multiple
blocks, it is crucial to establish consistent c-boundaries
for the physical variables across these blocks. In the
context of hydrodynamics, c-boundaries pertain only to the
interfaces between blocks, or block faces. The communication
patterns between these blocks are relatively simple and
direct, especially when there is no mesh refinement
involved. To further reduce the time spent on communication,
\kratos{} incorporates an additional optimization technique,
handling data transfers within the same device by directly
writing the ghost zone data to the destination, bypassing
the usual communication interfaces as described in section
\S\ref{sec:infra-comm}.


\kratos{} hydrodynamics relies on the the $2^d$-tree method
also for its mesh refinement, which plays a pivotal role in
managing the c-boundaries at the interfaces between refined
and coarser meshes. The general mechanism of c-boundary mesh
refinement is presented in Figure~\ref{fig:hydro_refine},
which illustrates the prolongation and restriction schemes
for primitive variables near the block boundaries that
separate blocks of different levels. The restriction
operations, which involve transferring data from finer to
coarser grids, are simply executed by performing
volume-weighted averages.  Conversely, the prolongation
step, which interpolates data from coarser to finer grids,
requires a more nuanced approach. This step must comply with
the TVD (total variance diminishing) condition to prevent
numerical instabilities that can arise from the
interpolation process. The minmod function,
\begin{equation}
  \label{eq:minmod}
  {\rm minmod}(a, b) =
  \begin{cases}
    & a\ ,\quad |a| < |b|\ {\rm and\ } ab > 0;\\
    & b\ ,\quad |a| \geq |b|\ {\rm and\ } ab > 0;\\
    & 0\ ,\quad ab < 0\ ,
  \end{cases}
\end{equation}
is used to evaluate the derivatives in the slope-limited
piecewise linear method (PLM). The minmod function is
crucial for maintaining the stability and accuracy of the
simulation by limiting the slope of the reconstructed
variables, thus preventing oscillations that could lead to
unphysical results.

The communication pattern in \kratos{} hydrodynamics is
designed in a specific sequence to ensure the consistency
and efficiency of computations:
\begin{enumerate}
\item Finishing all communication operations that are not
  directly related to cross-level block boundaries, to
  prepare for necessary data for prolongation;
\item Conducting prolongation calculations on the coarser
  blocks over all cross-level boundaries before sending the
  prolongated data to finer blocks, as the coarser blocks
  have complete information required;
\item Calculating the fluxes on all blocks (see
  \S\ref{sec:hydro-flux});
\item Sending the flux information at the cross-level block
  boundaries from the finer side to the coarser side, and
  sum up the fluxes accordingly to obtain the fluxes that
  are actually adopted on the coarser side at the
  cross-level interface, so that the conservation laws are
  maintained at the c-boundaries.
\end{enumerate}

\subsection{Flux calculation}
\label{sec:hydro-flux}

Almost all hydrodynamic methods in the FVM category have the
most time-consuming part on the computations of fluxes for
hydrodynamic variables. The efforts of improving the
performance of hydrodynamic simulations should focus on the
optimization of flux calculations.

\subsubsection{Shared memory utilization}
\label{sec:hydro-flux-shared}

The access of data in the global device memory
\footnote{Typically the memory spaces whose sizes marked
  explicitly by manufacturers as ``graphics memory''.}
(``global memory'' hereafter; not to be confused with the
system memory on the CPU side) suffer from excessive latency
(typically equivalent to $\sim 10^2$ or even more clock
cycles for the awating computing units) on hetrogeneous
devices. This issue becomes even worse for transposed data
access when calculating the fluxes along the directions
perpandicular to the $x$-axis. To reduce the cost of memory
access, almost all contemporary hetrogeneous computing
devices and their corresponding model have an important
feature, named ``shared memory'', that can significantly
improve the performance. Shared memory indicates memory
spaces that are typically allocated on the L1 cache of
stream multiprocessors (SMs). Such memory spaces are
equivalently fast as L1 when accessed by computing units,
while the data flow can be controlled by the programmer, in
contrast to the ordinary L1 cache spaces that are controlled
by the default dispatcher.

\begin{figure*}
  \centering
  \includegraphics[width=7.0in, keepaspectratio]
  {\figdir/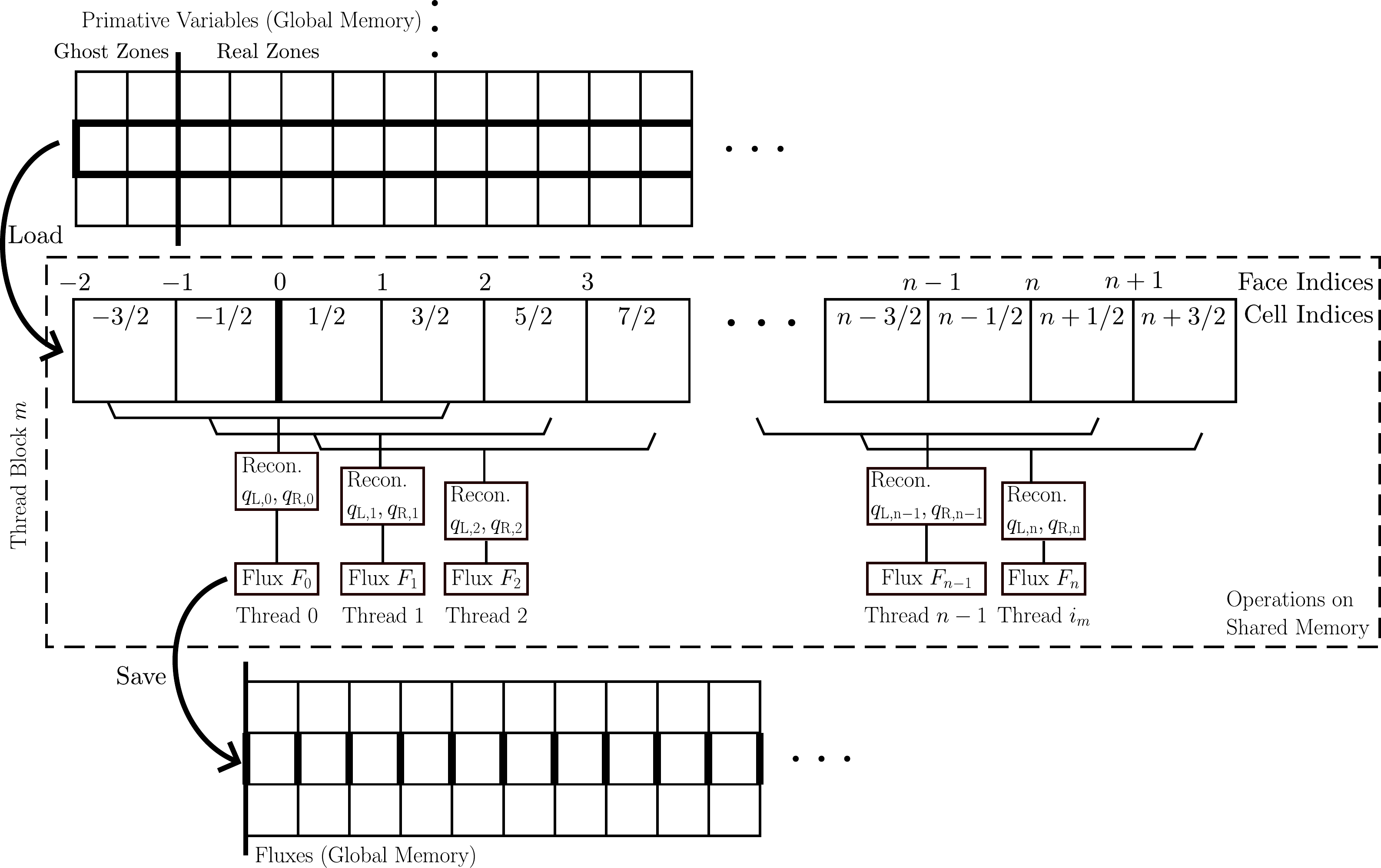}
  \caption{Patterns of calculating the fluxes for
    hydrodynamics, illustrating the utilization of shared
    memory. All processes within the middle dashed box use
    the shared memory instead of the global memory, as the
    memory-related operations are frequent, and the data in
    each cell is typically used many times by multiple
    threads (indicated by the braces and blocks in the
    middle box). }
  \label{fig:flux_calc_row} 
\end{figure*}

Proper utilization of the shared memory is a necessity for
maximizing the performance of flux calculations, minimizing
the frequency of accessing the data staying in the global
memory by adequately coordinating the data flow. The
reconstruction step, regardless of the actual algorithm,
obtains the physical variables on both sides of an interface
separating neighbouring cells. The $l$th order
reconstruction method will require data from $2(l+1)$ cells
for one interface, and in turn, data in one cell (not in
ghost zones) will be read by the reconstruction of $2(l+1)$
interfaces. These data dependencies, illustrted in
Figure~\ref{fig:flux_calc_row}, clearly indicates the
dispatching procedures for reconstruction, to load the row
of related data into the shared memory first, and then
conduct subsequent calculations using the data and space in
the shared memory.

\subsubsection{Mixed-precision methods}
\label{sec:hydro-flux-mixed}

In the \kratos{} framework, the hydrodynamics module is
implemented via a mixed-precision approach, a feature that
ensures both accuracy and computational efficiency. This
methodology defines a \texttt{float2\_t} data type (double
precision by default, typically represented by 64-bit
floating-point numbers), to maintain the conservation laws
of hydrodynamics with machine-level precision for
conservative variables, encompassing mass density, energy
density, and momentum density. Meanwhile, the hydrodynamics
module approximates the fluxes of these conservative
variables using approximate Riemann solvers, a common
practice in most of the finite volume hydrodynamics
simulation codes. Therefore, employing full double precision
methods for these calculations is not mandatory, provided
that numerical stability is guaranteed. To this end, the
hydrodynamics module opts for the \texttt{float\_t} data
type, which defaults to single precision (typically
represents 32-bit floating-point numbers), for the
reconstruction schemes and the approximate Riemann
solvers. This approach ensures that the conservation laws
are upheld to the precision inherent in \texttt{float2\_t},
while also maintaining physical reliability by holding the
conservation laws.

It is worth noting that, through the appropriate selection
of compilation options, the data types \texttt{float\_t} and
\texttt{float2\_t} can be mapped to either single or double
precision, offering flexibility in the precision
requirements of the simulation. This feature is useful in
scenarios where double precision is mandated, especially
when using devices that have comparable computational
capabilities for both single and double precision operations
(e.g. the Tesla series of computing GPUs).


The process of resolving fluxes is generally broken down
into two main steps. The initial step involves the
slope-limited PLM reconstruction, which is easily executed
in single precision. However, it is crucial to use the cell
center distance, $\Delta x$, directly in the denominator for
numerical derivatives to avoid significant errors that can
arise from large number subtractions, a phenomenon known as
``catastrophic cancellation''. The subsequent step involves
approximately solving the Riemann problem, a process similar
to the methods detailed in \cite{toro_riemann_2009}. The
HLLC solver is used as a default example, where the contact
surface speed $s_*$ is the crucial value, 
\begin{equation}
  \label{eq:hllc-contact-spd}
  s_* = \dfrac
  {[p_R - \rho_R u_R (s_R - u_R) ] -
    [p_L - \rho_L u_L (s_L - u_L) ] }
  {\rho_L(s_L-u_L) - \rho_R(s_R-u_R)}. 
\end{equation}
It is noted that, when $|s_i - u_i| \ll |u_i|$
($i\in \{L,R\}$) for either left (L) or right (R) states,
the reliability of the equation for $s_*$ is
compromised. Numerical tests have shown that single
precision calculations can lead to more instances of
catastrophic cancellations and unreliable results, or even
vanishing denominators in the equation for $s_*$. To
mitigate this, subtraction-safe modifications are
implemented,
\begin{equation}
  \label{eq:hllc-mixed-dvel}
  \begin{split}
    & \Delta s_L \equiv s_L - u_L = -c_{s,L} q_i\ ,\\
    & \Delta s_R \equiv s_R - u_R = c_{s,R} q_i\ ;\\
    & q_i = \left[ 1 + \dfrac{\gamma + 1}{2\gamma}
          \left(\max\left\{ \dfrac{p_*}{p_i}, 1\right\}
          - 1 \right) \right]^{1/2}.
  \end{split}
\end{equation}
leveraging the fact that $\Delta s_L$ is always negative and
$\Delta s_R$ is always positive, which prevents catastrophic
cancellations or vanishing denominators in
eq.~\eqref{eq:hllc-contact-spd}.

By applying a mixed-precision algorithm to the HLLC Riemann
solver within the \kratos{} module, extensive testing has
confirmed that the relative error remains below $10^{-3}$
when compared to the double precision version across the
parameter space relevant to typical astrophysical
hydrodynamic simulations (not shown in this paper). For
direct tests against analytic solutions in shock tube
scenarios as part of the Godunov solver pipeline, readers
are directed to \S\ref{sec:tests-shock-tube} for further
details.

\subsection{Time integrator}
\label{sec:hydro-intg}

The time integrator is an independent sub-module within the
hydrodynamic solver. By default, the time integration is
conducted by the Heun method, one of the widely adopted
second-order Runge-Kutta integrators, in which the
``one-step integration'' block is equivalent to the whole
cycle described in Figure~\ref{fig:hydro_proc}. One feature
of the Heun method is that the integrated variables of the
predictor sub-step (also known as the intermediate sub-step)
directly enters the results of the whole step with weight
$1/2$, which requires that the predictor sub-step results
should be recorded with the same precision as the whole
step.

\response{When the block refinement conditions are
  complicated, the \kratos{} system does not prescribe a
  universal time-stepping scheme. This hydrodynamic module
  adopts global time-stepping by default, while users are
  still allowed to include localized, block-specific time
  stepping schemes. In a wide variety of astrophysical
  scenarios, no more than $\sim 6$ levels of mesh refinement
  are typically adopted, including young stellar accretion
  disks, turbulent interstellar media, circumgalactic and
  intergalactic media, etc., where the global time-stepping
  scheme is already sufficient to support most scientific
  exploration goals. \kratos{} has also planned block-level
  adaptive time-stepping that is emerged into the
  mesh-refinement framework. Because the current framework
  already involves asynchronous communication
  (\S\ref{sec:infra-comm}) and load balancing based on space
  filling curves (\S\ref{sec:infra-module-manager}), proper
  weighting (which could be as simple as weight each block
  by the number of equivalent time steps on each global
  cycle) can resolve the issues associated with block-level
  adaptive time-stepping, and will be implemented when
  necessary.}

When the memory occupation is a concern, one can also select
the memory-optimized second order van Leer integrator.
This integrator can store the prediction sub-step results in
the space for primative varibales with reduced precision,
since the van Leer integrator does not1 directly use the
prediction step in the final results. Such approach saves up
to $\sim 1/3$ of the device memory for hydrodynamics, while
keeping the conservation law up to the full machine
accuracy.

\section{Verifications and performance tests}
\label{sec:tests}

Several fundamental and standard tests of the \kratos{} code
are presented in this section. Some of them are not only
verifying the correctness and robustness of the algorithms
involved, but also used as the opportunity of testing the
performance for the computing speed.

\subsection{Convergence tests}
\label{sec:tests-convergence}

\begin{figure}
  \centering
  \vspace*{-1.2cm}    
  \hspace*{-0.4cm}
  \includegraphics[width=3.8in, keepaspectratio]
  {\figdir/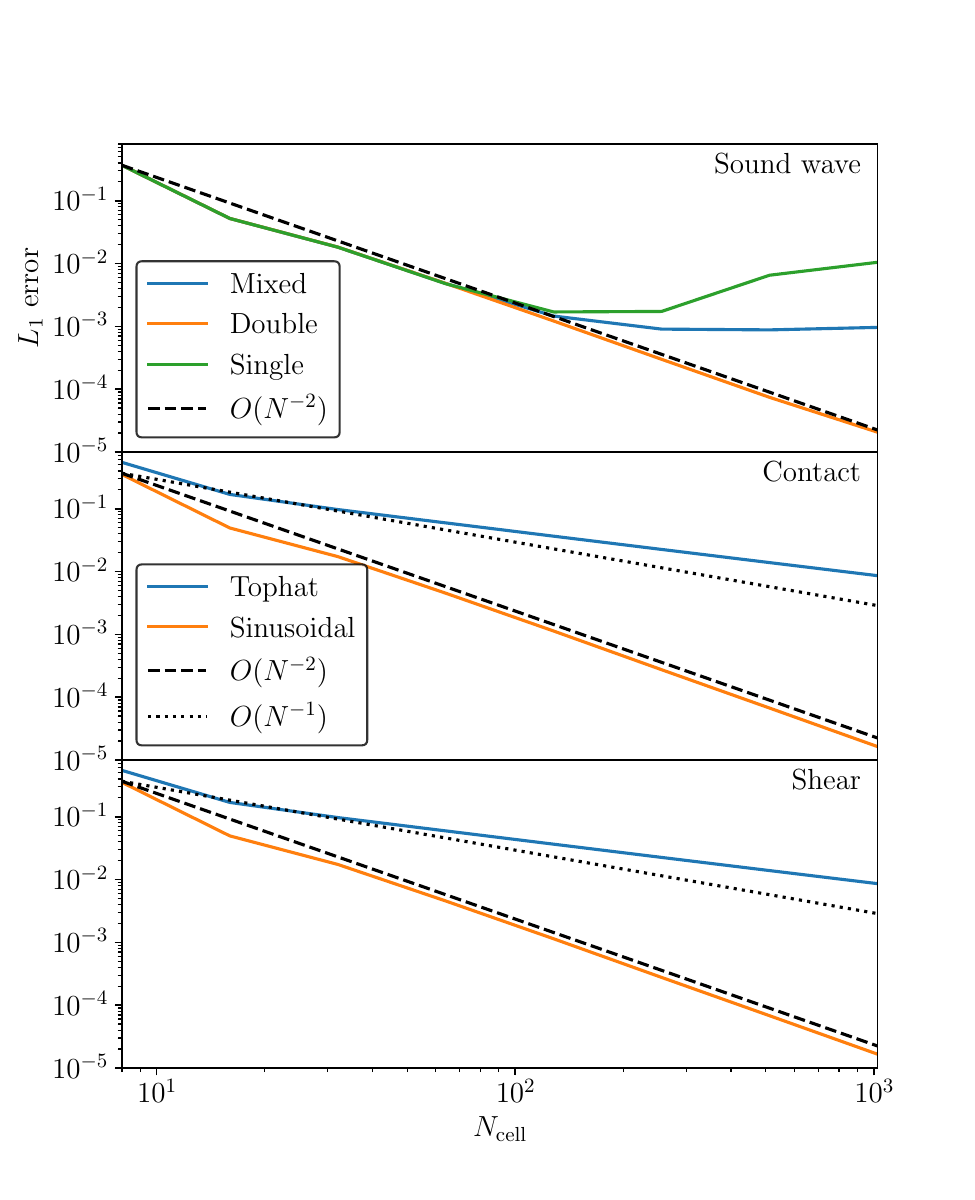}
  \caption{The $L_1$ errors in the convergence tests for the
    default hydrodynamic module in \kratos{} (see also
    \S\ref{sec:tests-convergence}). The top panel shows the
    sinusoidal sound wave tests, comparing the results with
    different computational precisions with the
    $L_1\propto N^{-2}$ power-law. The lower two panels
    exhibit the results for contact waves and shear waves
    with tophat and sinusoidal initial conditions,
    respectively, using the mixed precision methods (see
    also \S\ref{sec:hydro-flux-mixed}; note that full double
    precision results are almost identical to the mixed
    precision ones, and are thus not illustrated).  }
  \label{fig:test_convergence}
\end{figure}

\response{ The fundamental tests of hydrodynamic methods and
  algorithms involve evolving the eigenfunctions of
  hydrodynamic equations, which include sound waves, contact
  waves, and shear waves. When periodic boundary conditions
  are applied, these eigenfunctions should ideally revert to
  their initial conditions after one kinematic crossing
  time, defined as $t_{\rm cross} \equiv l_{\rm box} / v$,
  where $l_{\rm box}$ represents the size of simulation
  domain (box), and $v$ denotes the speed of the eigenmode
  along the specific direction of motion.  To quantify the
  difference between the initial and evolved states, the
  normalized $L_1$ error is introduced as,
  \begin{equation}
    \label{eq:l1_err}
    L_1[q] \equiv \dfrac{\sum_i \left| q_i - q_i^0 \right|}
    {N_{\rm cell} \left[ \max\{q_i^0\} - \min\{q_i^0\}
      \right] }\ , 
  \end{equation}
  where $\{q_i^0\}$ and $\{q_i\}$ correspond to the values
  of physical quantity $q$ on the mesh before and after
  evolution, respectively. For clarity, the tests are
  initialized with background physical quantities
  $\rho_0 = 1$, $p_0 = 5/3$, and $\gamma = 5/3$. The
  one-dimensional simulation domain, which has periodic
  boundaries, is set to have a size of $l_{\rm box} = 1$.
  Different types of waves are superimposed as perturbations
  $\{\delta q_i\}$ on this domain, using either a sinusoidal
  function [$\delta q_i = A \sin(2\pi x_i)$], or a tophat
  function ($\delta q_i = A$ if $x_i < 0.5$ and
  $\delta q_i = 0$ elsewhere). All tests are run from
  $t = 0$ to $t = 1$, covering a full period of
  motion. Tests with $v=-1$ are also carried out, generating
  identical results to confirm the reflection symmetry of
  the algorithms adopted.

  The test results are illustrated in
  Figure~\ref{fig:test_convergence}. For the sound wave
  tests, it should be noted that the linearized sinusoidal
  sound wave serves as an approximate eigenfunction of the
  hydrodynamic equations. Such approximation necessitates
  sufficiently small amplitudes (numerical experiments
  indicate that $A\lesssim 10^{-6}$ is required) to ensure
  the wave evolving without physical distortions. However,
  when the amplitude is small and the resolution is high,
  the small contrasts between two adjacent cells
  significantly reduce the accuracy of reconstruction, which
  undermines the proper shapes of waves and the subsequent
  measurements of the $L_1$ errors. Consequently, for sound
  wave tests employing mixed and full single-precision
  methods, the amplitude is set at $A=10^{-4}$, which
  prevents the $L_1$ errors from decreasing further with
  resolutions higher than $N_{\rm cell}\simeq 10^2$. In
  contrast, the full double precision case uses
  $A = 10^{-6}$ and achieves convergence up to
  $N_{\rm cell} > 10^3$, following a scaling rule of
  $L_1\proptosim N_{\rm cell}^{-2} $.  For contact waves
  (where only mass density varies) and shear waves (where
  only tangential velocity varies), a relatively larger
  amplitude ($A = 10^{-1}$) can be adopted, and the $L_1$
  errors of mixed-precision methods are almost identical to
  the full double-precision methods (omitted in this
  paper). Although the tophat initial condition exhibits
  slower convergence than $L_1\propto N_{\rm cell}^{-1}$,
  the sinusoidal waves still converge as
  $L_1\proptosim N_{\rm cell}^{-2}$.  This convergence
  behavior aligns with that of other existing numerical
  simulation codes utilizing the same hydrodynamic methods
  and algorithms \citep[e.g.,][]{2008ApJS..178..137S,
    stone_athena_2020, 2024arXiv240916053S}.
  }

\subsection{Sod shock tubes}
\label{sec:tests-shock-tube}

\begin{figure*}
  \centering
  \hspace*{-0.4cm}
  \includegraphics[width=7.5in, keepaspectratio]
  {\figdir/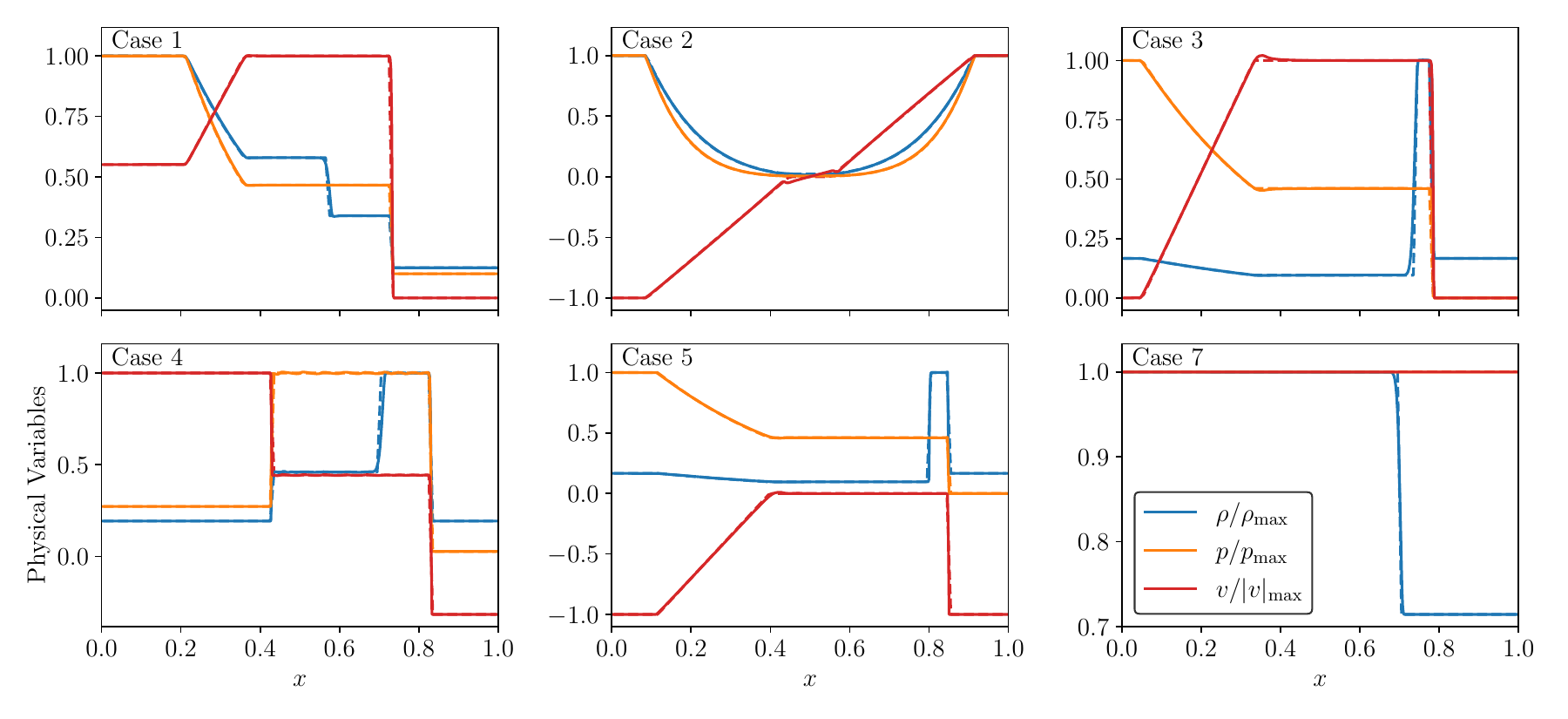}
  \caption{The one-dimensional tests of Sod shock tubes,
    showing the results of cases 1--5 and 7 in
    \citet{toro_riemann_2009} using the original setups and
    parameters. The exact semi-analytic solutions are
    indicated by dashed lines, in the same color for the
    corresponding physical quantities of \kratos{} numerical
    calculation results. All calculations in these tests use
    mixed precision methods.}
  \label{fig:test_sod_toro} 
\end{figure*}

The shock tube problem serves as a canonical benchmark for
assessing the accuracy and robustness of hydrodynamic
solvers, as well as the integrity of the overall simulation
infrastructure. While it is acknowledged that the presence
of expansion fans within the shock tube scenario precludes
the existence of exact analytic solutions, the derivation of
semi-analytic solutions remains a feasible and
straightforward endeavor. In this context,  a
series of simulation results are presented and depicted in
Figure~\ref{fig:test_sod_toro}, which correspond to a
variety of initial configurations. These configurations are
drawn from the standard test problems outlined in
\citep{toro_riemann_2009}. It is important to highlight that
the mixed precision HLLC approximate Riemann solver has been
deliberately selected for these tests, yielding results that
demonstrate a consistent alignment with the analytic
solutions across all examined cases.

Furthermore, extended tests based on the standard Sod shock
tubes are also conducted, which traditionally consider fluid
motion along the $x$-axis, by conducting additional tests
that interchange the axes of fluid motion. The findings from
these modified tests are found to be identical to those
obtained from the standard x-axis tests. This congruence
underscores the versatility of the finite volume scheme
employed, which facilitates the calculation of fluxes
through the pre-update primitive variables, thereby
highlighting one of the key advantages of this approach.

\subsection{Double Mach reflection test}
\label{sec:tests-dm-ref}

\begin{figure} 
  \centering
  \hspace*{-0.9cm}
  \includegraphics[width=4.in, keepaspectratio]
  {\figdir/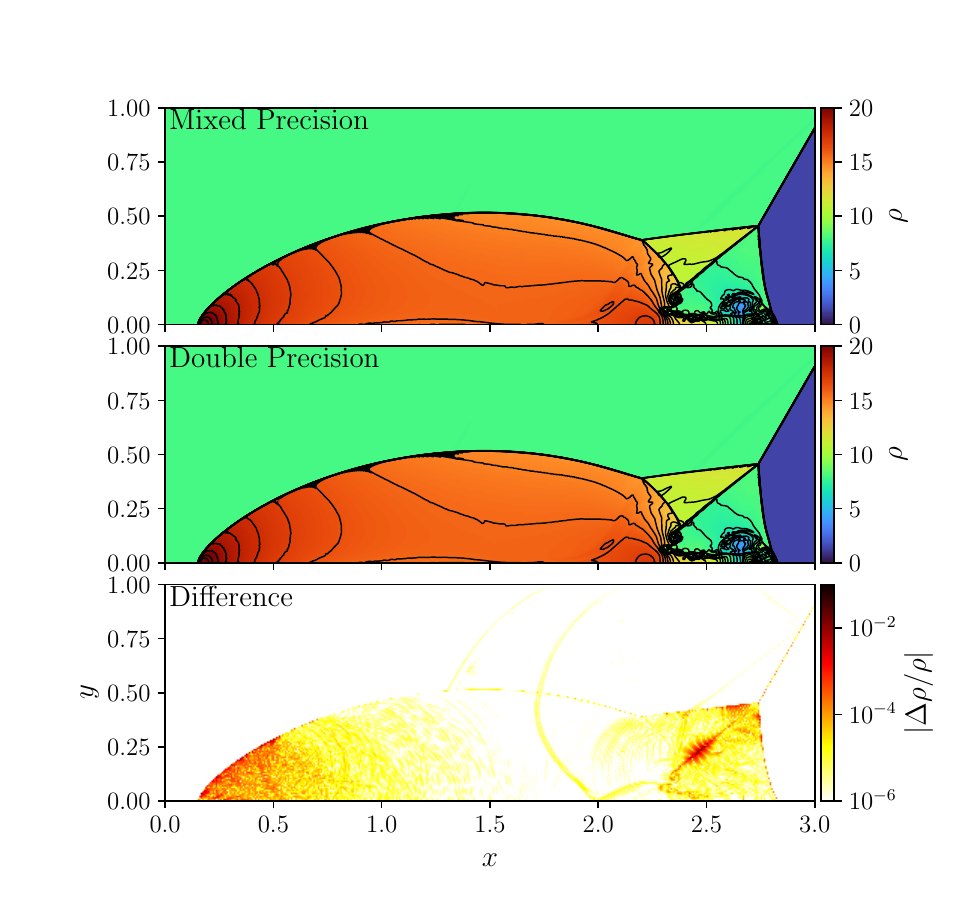}
  \caption{Results of double Mach reflection tests, using
    mixed precsision (top panel) and full double precision
    (middle panel) methods. The 30 contour lines starting
    from $\rho = 1.73$ to $\rho = 2.1$ are spaced evenly in
    $\rho$, indicating almost identical patterns on these
    panels. The bottom panel illustrate the relative
    differences of mass density comparing the top two
    panels. }
  \label{fig:test_double_mach} 
\end{figure}

The double Mach reflection tests, as canonical benchmarks
for hydrodynamics simulations, is employed to rigorously
evaluate the precision and dependability of simulation
frameworks. This test, initially formulated by
\citet{1984JCoPh..54..115W}, presents a scenario where a
Mach 10 shock wave impinges upon an inclined plane, leading
to complex interactions with the reflected shock wave and
the emergence of a triple point. These dynamics yield a
spectrum of hydrodynamic phenomena, including shock
discontinuities and the generation of a high-velocity
jet. The intricate flow structures that result are
instrumental in assessing a code's capacity to accurately
and robustly manage multi-dimensional discontinuities,
particularly shock waves. For this test case, the initial
conditions and boundary specifications are derived directly
from \citet{1984JCoPh..54..115W}, with a slightly expanded
spatial domain of $(x,y) \in [0,4] \times [0,1]$ and a
resolution of $4096 \times 1024$. To accurately simulate the
propagation of the inclined incident shock, a time-dependent
boundary condition is implemented along the upper boundary,
leveraging the p-boundary condition framework detailed in
Section \ref{sec:hydro-p-bnd}.

Figure \ref{fig:test_double_mach} illustrates contour plots
superimposed on a colormap representation of the solution at
$t = 0.2$, juxtaposing mixed and full-double precision
methods under identical conditions. It is evident that the
two algorithms, despite their differing precision levels,
exhibit near-perfect convergence, with the maximum relative
discrepancy not exceeding $\sim 10^{-2}$ as indicated by the
mass density. These outcomes closely mirror Figure 4 in
\citet{1984JCoPh..54..115W}, with the distinction that our
results showcase sharper shock fronts attributable to
higher-order spatial accuracy and a significantly enhanced
resolution. In comparison with the analyses presented in
\citet{2008CS&D....1a5005G} and \citet{stone_athena_2020},
the carbuncle-like instability near the jet's head is not
entirely mitigated by the HLLC solver, which is known to
introduce less numerical diffusion than the HLLE
solver. Nevertheless, the near-identical behavior between
mixed and double precision results at the jet head further
substantiates the validity and robustness of the mixed
precision solver.

The performance assessments based on the double Mach
reflection problem are summarized in
Table~\ref{table:spd-dmach}, with $10^8-10^9$ cells per
second computational efficiency {\it per device} on
contemporary GPUs. On GPUs for the consumer market (such as
the NVIDIA RTX series and AMD 7900XTX), mixed precision
calculations run at speeds comparable to the full single
precision (around $\sim 70-80\%$), and are almost
$\gtrsim 5-7$ times faster than the full double precision
runs. These speeds are not proportional to the theoretical
single and double precision floating point computing
efficiency of GPUs, as deeper profiling and timing tests of
\kratos{} reveal that, even with relatively thorough
optimization on the data exchange and cache or shared memory
utilization (\S\ref{sec:hydro-flux-shared}), a significant
fraction of computing time is still occupied by the data
exchange between the global graphics memory and the cache
(not shown in this paper). On computing GPUs (such as Tesla
A100 and MI100) with enhanced double precision performance,
the mixed precision methods do not exhibit considerable
advantages compared to full double precision. The
compatibility of \kratos{} on x64 and ARM CPUs is also
confirmed and illustrated in Table~\ref{table:spd-dmach}.
It is noted that the performance with HIP-CPU per physical
CPU core are far below the other CPU codes on same devices,
by approximately $\sim 3-7$ times
\citep[e.g.][]{stone_athena_2020}. Because the HIP-CPU
actually emulates GPUs using non-optimized patterns
(including the implementation of multi-threading dispatching
algorithms and launching overheads), \kratos{} on HIP-CPU is
implemented primarily for compatibility and code devloping
(especially debugging) considerations, rather performance
purposes.

\begin{deluxetable}{lccc}
  \tablecolumns{4} 
  \tabletypesize{\scriptsize}
  \tablecaption{Performance test results of double Mach
    reflections.
    \label{table:spd-dmach}
  } \tablehead{ \colhead{Programming Models} &
    \multicolumn{3}{c}{Computing Speed with Precision}
    \\
    \colhead{and Devices} &
    \multicolumn{3}{c}{($10^8{\rm\ cell\ s}^{-1}$)}
    \\
    \cline{2-4} \colhead{} & \colhead{Single} &
    \colhead{Mixed} & \colhead{Double} } \startdata
  HIP-CPU${}^*$ & & & \\
  AMD Ryzen 5950X${}^\dagger$ & & & 0.15 \\
  Qualcomm Snapdragon 888${}^{**}$ & & & $0.0037$ \\
  \\
  NVIDIA CUDA & & & \\
  RTX 2080TI & 9.2 & 6.9 & 0.93 \\
  RTX 3090   & 14  & 10  & 1.4 \\
  RTX 4090   & 21  & 16  & 2.9 \\
  Tesla A100 & 16 & 14  & 8.3 \\
  \\
  AMD HIP & & & \\
  7900XTX & 8.9 & 7.7 & 2.8 \\
  MI100   & 8.2 & 6.8 & 3.0 \\
  \enddata \tablecomments{ Presenting the average over
    $10^2$ steps. All tests cases are in 2D. Detailed setups
    see \S\ref{sec:tests-dm-ref}.  \\
    *: Only double precision results are concerned, as
    modern CPUs have almost the same single
    and double precision computing speeds.  \\
    $\dagger$: Utilizing all 16  physical cores.\\
    **: Using termux (\url{https://termux.dev}) on Android
    operating system, utilizing one major physical
    core. Compile-time optimization are turned {\it off}
    because of the software restrictions of \code{TBB} on
    ARM CPUs.  }
\end{deluxetable}

\subsection{Liska-Wendroff implosions }
\label{sec:tests-lw-implode}

\begin{figure*}
  \centering
  \hspace*{-0.4cm}
  \vspace*{-0.2cm}  
  \includegraphics[width=6.5in, keepaspectratio]
  {\figdir/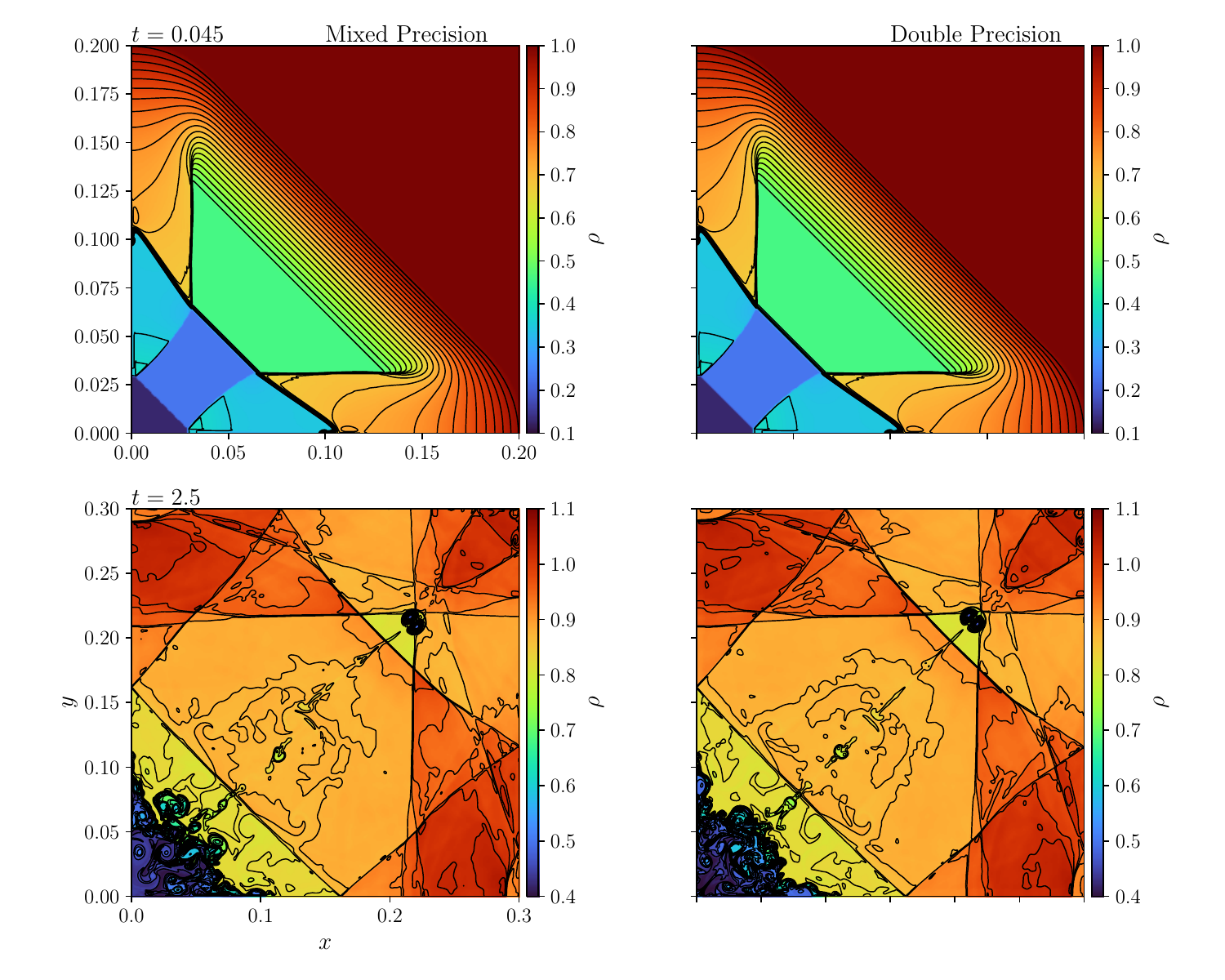}
  \vspace*{-0.5cm}
  \caption{The Liska-Wendroff implosion tests results,
    showing $t= 0.045$ (upper row) and $t = 2.5$ (lower
    row), comparing the mixed precision (left column) and
    full double precision (right column) methods. Contours
    span from $\rho = 0.35$ to $\rho = 1.1$, spaced evenly
    by $\delta \rho = 0.025$.  }
  \label{fig:test_lw} 
\end{figure*}

The Liska-Wendroff implosion test, as detailed in
\citet{LiskaWendroff2003}, serves as a benchmark for
assessing the directional symmetry preservation of
hydrodynamic simulation methods. This test is initiated with
an initial condition that is similar to the Sod shock-tube
scenarios, but with a diagonal orientation. A shock wave in
the $\gamma = 1.4$ gas is generated by applying an
under-pressure region in the lower-left corner of the
domain,
\begin{equation}
  \label{eq:lw-init}
  (\rho, p, \mathbf{v} ) =
  \begin{cases}
    & (0.125, 0.14, 0)\ ,\quad x+y < 0.15\ ;\\
    & (1, 1, 0)\ ,\quad \text{elsewhere}\ ,
  \end{cases}
\end{equation}
which is then reflected by the bottom and left
boundaries. The initial discontinuity is set at a $\pi/4$
angle to the coordinate axes, with reflecting boundary
conditions applied on all sides. The reflected shocks
interact with other discontinuities, including contact
discontinuities and additional shocks, leading to the
development of finger-like structures via the
Richtmeyer-Meshkov instability along the direction of the
initial normal vector. The direction of the jet serves as a
critical indicator of whether the simulation method
maintains reflective symmetry to machine precision.  In some
early hydrodynamic simulation systems that employed
directionally split algorithms, the jet's impact at the
lower-left corner might not align precisely with the corner,
resulting in vortices that deviate significantly from the
diagonal line. This deviation highlights the importance of
directional symmetry in accurately capturing the dynamics of
the system.

Figure~\ref{fig:test_lw} illustrates the density profiles at
$t=0.045$ and $t=2.5$, comparing the results obtained from
mixed precision and double precision methods. The double
precision results exhibit good symmetry across the diagonal
at $t=2.5$, with the exception of the lower left corner,
which is prone to strong instabilities and turbulence. The
mixed precision results also maintain a large degree of
reflection symmetry; however, the wake of the jet-induced
vortices shows a noticeable departure from perfect symmetry
when visualized with contour plots. It is important to note
that the last significant digits in floating-point
calculations on GPUs are not guaranteed to be identical for
both single and double precision, and the associated errors
are inherently undetermined
\citep[e.g.][]{gpu_ieee}. Amplified by chaotic motions, such
random errors eventually lead to the symmetry breaking.
Before employing mixed precision methods in practical
applications, it is crucial to ascertain whether strict
reflection symmetry is necessary. Additionally, it should be
recognized that this type of discrete symmetry breaking,
caused by slightly differing results from separate
calculations, does not affect continuous symmetries, such as
the conservation laws of mass and momentum.

\subsection{Kelvin-Helmhotz instabilities}
\label{sec:tests-khi}

\begin{figure*}
  \centering
  \hspace*{-0.6cm} 
  \includegraphics[width=7.5in, keepaspectratio]
  {\figdir/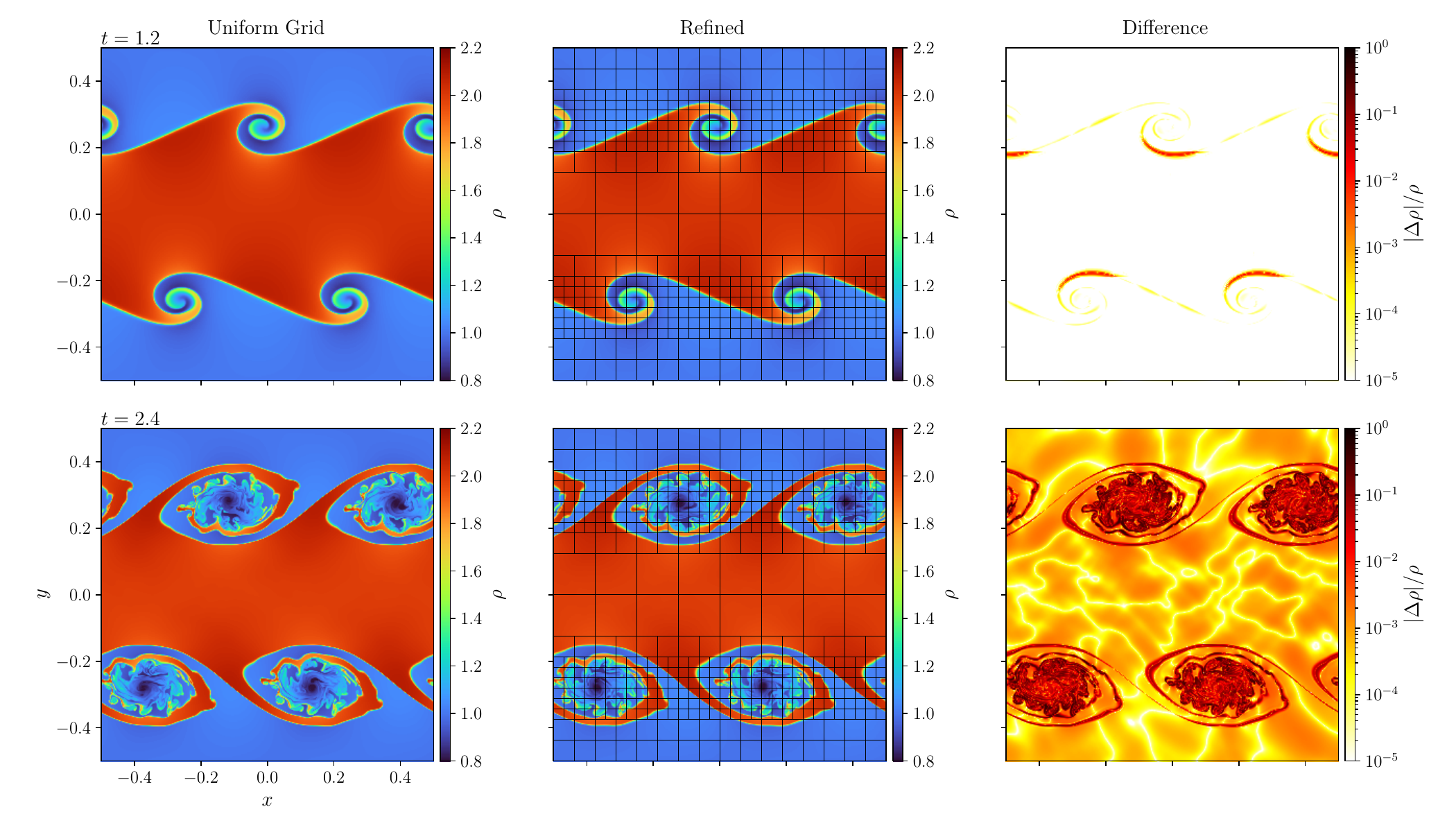}
  \caption{The Kelvin-Helmhotz instability tests results,
    comparing the $t=1.2$ (upper row) and $t=2.4$ (lower
    row) for the uniform grid (left column) and SMR (middle
    column) results for the relative difference (right
    panel) in mass density. The block boundaries are
    indicated in the panels for the SMR simulation, showing
    the mesh refinement results (note that each block has
    the same number of cells).  }
  \label{fig:test_kh}
\end{figure*}

The tests on Kelvin-Helmhotz instabilities are one of the
``standard procedures'' for many numerical simulation
systems. Such tests, similar to \citet{2016MNRAS.455.4274L},
are conducted on $\gamma = 1.4$ gases with initial
conditions,
\begin{equation}
  \label{eq:kh-init}
  \begin{split}
    & \rho = \rho_0 + \Delta \rho \tanh
      \left( \dfrac{|y - y_0|}{L} \right)\ , \\
    & v_x = v_{x0} \tanh \left( \dfrac{|y - y_0|}{L}
      \right)\ ,\\
    & v_y = v_{y0} \cos(k x) \exp \left[ -\dfrac{(y -
      y_0)^2}{2\sigma^2} \right]\ ,
  \end{split}
\end{equation}
where $\rho_0$, $v_{x0}$ and $v_{y0}$ are the reference
values for density and $x, y$ velocity components,
$\Delta \rho$ is the difference in density across the
shearing layer with thickness $L$, $k$ is the velocity
perturbation wavenumber, and $\sigma$ is the thickness of
the perturbation region. This paper chooses the parameters
$L = 0.01$, $y_0 = 0.25$, $\rho = 1.5$, $\Delta \rho = 0.5$,
$v_{x0} = 0.5$, $v_y = 0.01$, $k = 4\pi$, and
$\sigma = 0.2$. These 2D simulations are carried out within
a spatial domain of
${x,y} \in [-0.5, 0.5] \times [-0.5, 0.5]$. The uniform grid
simulations employ a resolution of $\Delta x = 1/2048$
across the entire domain, whereas the mesh refinement
approach ensures this resolution is maintained in the
regions $|x| \in (3/16, 3/8)$ through the application of two
levels of static mesh refinement. Periodic boundary
conditions are applied to the $x$ boundaries, and reflective
conditions elsewhere. 

The results of these tests are illustrated in
Figure~\ref{fig:test_kh}, which compares the outcomes from
uniform grid simulations with those from simulations
utilizing mesh refinement. At $t = 1.2$, the upper row of
the figure illustrates results that are nearly
indistinguishable between the two methods, with a relative
error not exceeding $5 \times 10^{-2}$. The minor
differences observed are primarily localized near regions of
density discontinuities, likely due to subtle structural
displacements. At $t = 2.4$, the patterns remain largely
consistent, yet the quantitative differences between the
uniform grid and mesh refinement results are more
pronounced, reaching up to $\sim 10^{-1}$. These
discrepancies are also attributed to structural
displacements, as evidenced by a direct comparison of the
uniform grid and mesh refinement panels. It is important to
note that simulations employing mesh refinement may still
exhibit variations from those with high-resolution uniform
grids, particularly in the damping of short-wavelength modes
within unrefined, low-resolution areas. The kinetic
diffusivity parameter, which is influenced by numerical
methods, scales with $\eta \propto \Delta x^2/\Delta
t$. This phenomenon has also been observed in various
studies, and is further supported by the work of
\citet{stone_athena_2020}.

\subsection{Rayleigh-Taylor instabilities}
\label{sec:tests-rti}

\begin{figure}
  \centering
  \hspace*{-0.4cm}
  \includegraphics[width=3.5in, keepaspectratio]
  {\figdir/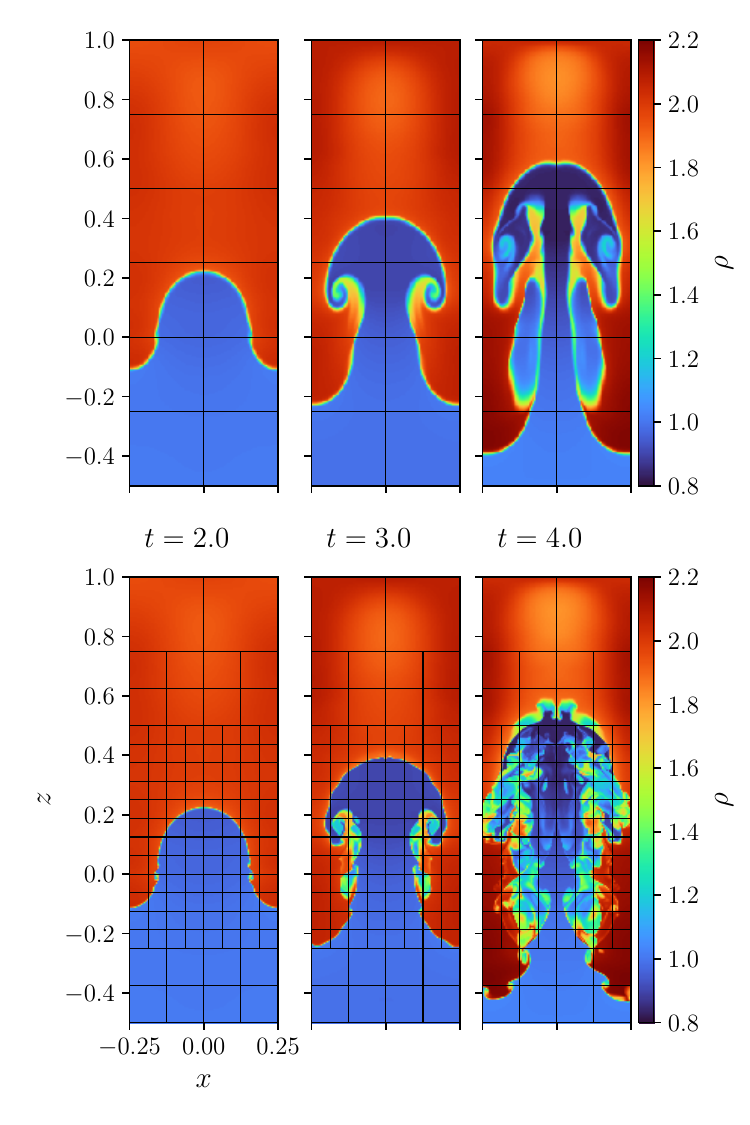}
  \vspace*{-1.0cm}  
  \caption{The Rayleigh-Taylor instability tests, showing
    $t = 2, 3$ and $4$ results in three columns for the
    simulations on uniform grid (top row) and with SMR
    (bottom row). Similar to Figure~\ref{fig:test_kh}, the
    block boundaries are indicated to illustrate the mesh
    refinement layouts.}
  \label{fig:test_rt} 
\end{figure}

The Rayleigh-Taylor instability (RTI) test, a standard
benchmark for evaluating the efficacy of numerical methods
in multi-dimensional hydrodynamics, is further detailed in
e.g. \citet{LiskaWendroff2003} and
\citet{2016MNRAS.455.4274L}. This test is useful for
assessing how numerical schemes behave in the presence of
complex fluid interactions. This test also examines the
performance of the code under static mesh refinement, as the
regions where significant hydrodynamic features are
anticipated to emerge can be identified with relative ease
{\it a priori}.

The simulated domain covers
$(x, y, z) \in [-1/4,1/4] \otimes [-1/4,1/4] \otimes
[-1/2,1]$ with $\gamma = 1.4$ gas. While the uniform grid
test employs a grid spacing of $\Delta x = 1/256$, the mesh
refinement scenario focuses on a refined region of
$(x, y, z) \in [-1/4,1/4] \otimes [-1/4,1/4] \otimes
[-1/2,1]$ to the second level, achieving a grid spacing of
$\Delta x = 1/1024$. Reflecting boundary conditions are
implemented on all physical boundaries. Initially, the
density is set to $\rho = 2$ for $z > 0$ and $\rho = 1$ for
$z \leq 0$. The pressure is determined based on the vertical
gravitational acceleration $g_z = -0.5$ (negative indicating
the downward direction of gravity), ensuring a hydrostatic
equilibrium calibrated at $p_{z=0} = 2.5$. An initial
velocity perturbation is introduced as
$v_z = 10^{-2} \cos(\mathbf{k} \cdot \mathbf{x})$, where
$\mathbf{k} \equiv (4\pi, 4\pi, 3\pi)$ represents the
perturbation wavenumber.

The outcomes of the tests, encompassing mesh refinement
conditions, are depicted in Figure~\ref{fig:test_rt},
presenting a comparative analysis between simulations with
and without mesh refinement at various evolutionary
times. It is important to note that, due to the absence of
explicit viscosity or surface tension, the emergent shear
instabilities parasitic on the Rayleigh-Taylor instability
patterns are theoretically expected to grow at nearly all
wavelengths. The actual results are, therefore, affected by
the numerical diffusivities, which are proportional to
$\eta \propto \Delta x^2/\Delta t$. This leads to
significant differences when comparing mesh refinement
results with those obtained using low-resolution uniform
grids. With equivalent resolution of $\Delta x = 1/1024$ on
the refined mesh, the non-linear fine structures are more
pronounced due to reduced numerical diffusivities, while the
overall ``mushroom'' structures remain quantitatively
similar for both scenarios.

The performance metrics of \kratos{} in these tests are
summarized in Table~\ref{table:spd-rt}. As the
Rayleigh-Taylor instability tests are conducted in 3D, the
time consumption due to the flux calculations on the third
dimension nominall reduces the speed (in cells per second)
by around $\sim 1/3$ when compared to the 2D results in
Table~\ref{table:spd-dmach}. One can also observe by
comparing the single-device speed with dual-device speed
(the first and second numbers of each data entry in the
table) that the cross-device parallelization performance is
good with two devices other than RTX 4090 (up to $\sim 90\%$
the theoretical performance). This indicates that the
workflow design (see also \S\ref{sec:hydro} and
Figure~\ref{fig:hydro_proc}) is successfully optimizing out
the communication overheads. When using the RTX 4090 GPUs,
however, the scalability appears to be worse, as the high
single-GPU performance partially fails the workflow in
hiding the communication operations behind computations.
\response{It is noted that modern CPU-based codes carry out
  hydrodynamic simulations with the same temporal and
  spatial accuracy (also using the slope-limited PLM
  reconstruction, the HLLC Riemann solver, and the
  second-order Heun integrator) and same mesh refinement
  configurations at a speed of
  $\sim 3\times 10^6~{\rm cell\ s}^{-1}$ per CPU core (e.g.,
  Athena++; see \citealt{stone_athena_2020}), or
  $\sim 3\times 10^8~{\rm cell\ s}^{-1}$ per 128-core
  computing node given perfect scalability. This speed is
  equivalent to $\sim 30\%$ of one RTX 4090 GPU, and one may
  have noticed that each node can be equipped with multiple
  (up to 8) GPUs, leading to $\sim 20$ times computing
  performance improvement per computing node.  }

\begin{figure*}
  \hspace*{-0.7cm}
  \includegraphics[width=7.5in, keepaspectratio]
  {\figdir/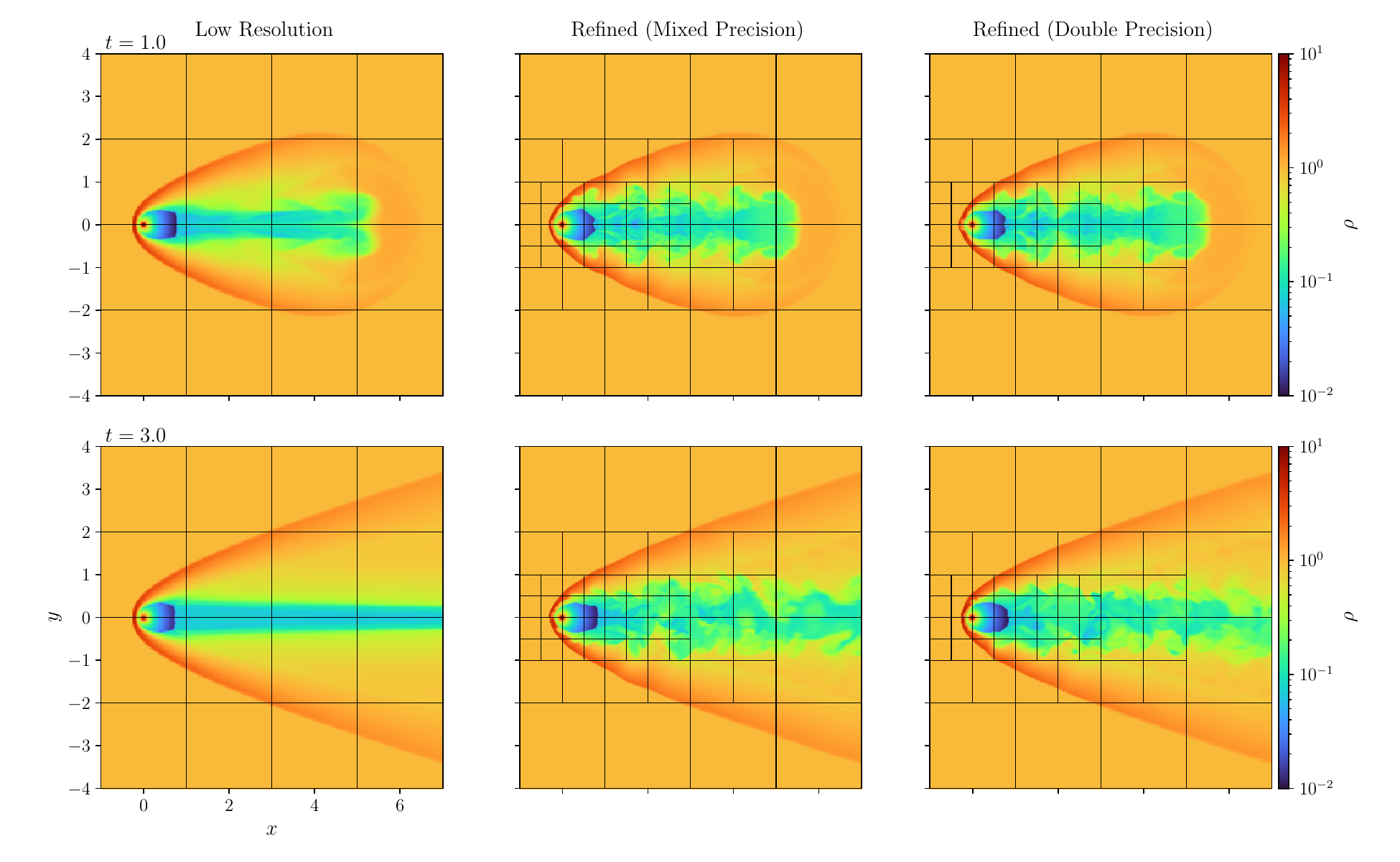}\vspace*{-0.8cm}\\
  \includegraphics[width=2.24in, keepaspectratio]
  {\figdir/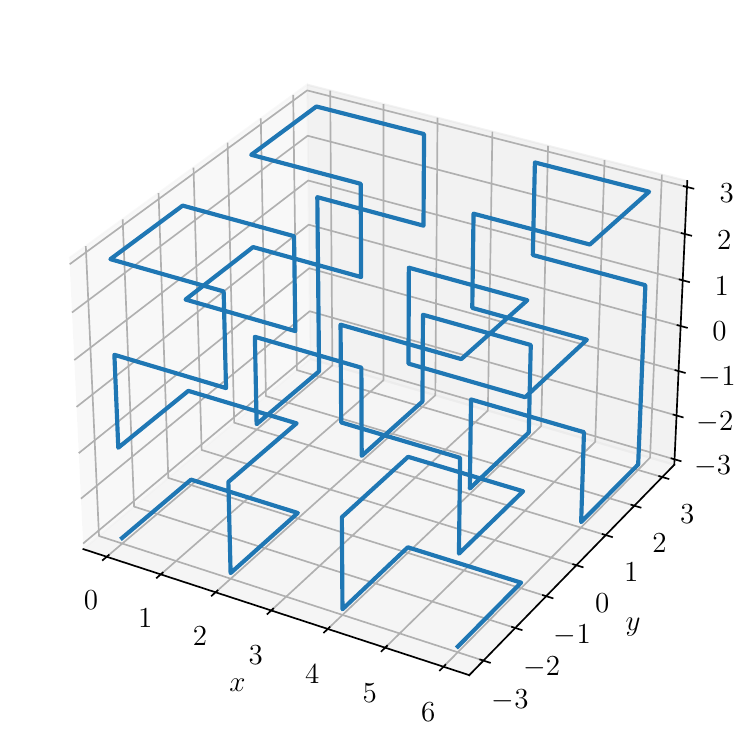}
  \includegraphics[width=2.24in, keepaspectratio]
  {\figdir/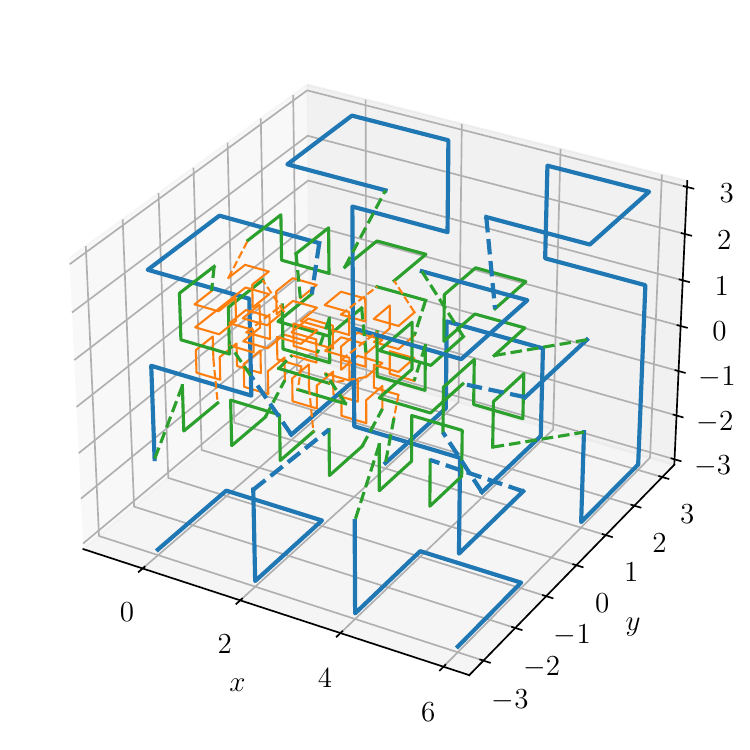}
  \includegraphics[width=2.2in, keepaspectratio]
  {\figdir/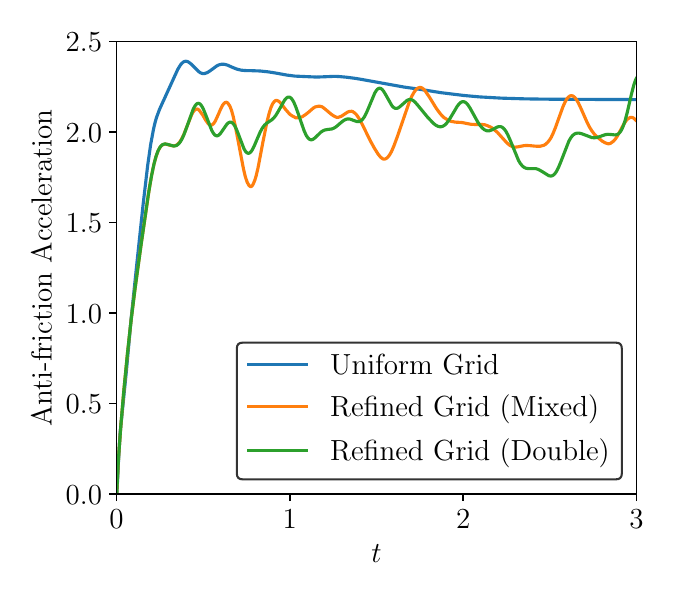}  
  \caption{ Results of the tests on colliding outflows and
    ambient gases, showing the mas density profiles at
    $t = 1$ (top row) and $t = 2$ (middle row) for low
    resolution ($\Delta x = 1/32$; left column), SMR with
    mixed precision (equivalent $\Delta x = 128$; middle
    column) and SMR with full double precsion (right column)
    cases. Block boundaries are indicated in related panels
    to illustrate the mesh layouts. The bottom row
    illustrates the Hilbert curves generated for the job
    dispatching and load balancing for the uniform grid
    (left panel) and SMR (middle panel) simulations, where
    the lines connecting blocks on different levels are
    shown in different colors, and cross-level connections
    are indicated by dashed lines. The evolution history for
    the acceleration on the central object of the three
    different simulations (lower right panel).}
  \label{fig:test_bow_shock} 
\end{figure*}

\begin{deluxetable}{lccc}
  \tablecolumns{4} 
  \tabletypesize{\scriptsize}
  \tablecaption{Performance test results based on
    the Rayleigh-Taylor instability simulations.
    \label{table:spd-rt}}
  \tablehead{ \colhead{Programming Models} &
    \multicolumn{3}{c}{Computing Speed with Precision}
    \\
    \colhead{and Devices${}^\dagger$} &
    \multicolumn{3}{c}{($10^8{\rm\ cell\ s}^{-1}$)}
    \\
    \cline{2-4} \colhead{} & \colhead{Single} &
    \colhead{Mixed} & \colhead{Double} } \startdata
  HIP-CPU & & & \\
  AMD Ryzen 5950X${}^*$ & & & (0.15, - ) \\
  \\
  NVIDIA CUDA & & & \\
  RTX 3090   & (8.6, 14.7)  & (6.2, 11.2)  & (0.61, 1.1) \\
  RTX 4090   & (12.4, 17.7)  & (10.1, 16.3)  & (1.3, 2.4) \\
  Tesla A100 & (8.3, 14.6) & (8.3, 12.9)  & (4.7, 8.2) \\
  \\
  AMD HIP & & & \\
  7900XTX & (3.6, 5.7) & (3.4, 5.5) & (0.94, 1.86) \\
  MI100${}^\ddagger$ & (3.7, -) & (3.6, -) & (2.1, -) \\
  \enddata \tablecomments{ Presenting the average over
    $10^2$ steps. Results are presented as groups of numbers
    ($s_1, s_2$), where $s_1$ is for the single-devicle
    speed, and $s_2$ for the dual-device one.  All tests
    cases are in 3D (see \S\ref{sec:tests-rti}). \\
    $\dagger$: Some devices involved in
    Table~\ref{table:spd-dmach} are not applicable in
    these tests.\\
    *: Dual-device tests not applicable.\\
    $\ddagger$: Tests are not conducted on dual-device MI100
    systems due to technical restrictions. }
\end{deluxetable}

\subsection{Collidingas outflow and ambients}
\label{sec:tests-outflow-amb}
 
A traditional yet highly relevant scenario, which could
showcase the performance and capabilities of the simulation,
involves interactions between stellar outflows and ambient
gas flows induced by the relative motion of stars and their
surrounding medium. This scenario is not only of historical
significance but also gains contemporary relevance in the
dynamical studies of stars and compact objects, as
highlighted in recent literatures \citep[e.g.][]
{2020MNRAS.494.2327L, 2022ApJ...932..108W}.

These tests are conducted for $\gamma = 1.4$ gas, within a
cubic spatial domain defined by the coordinates
$(x, y, z) \in [-1,7] \otimes [-4,4] \otimes [-4,4]$. At the
left boundary along the $x$-axis, an inflow is introduced
with a velocity of $v_{\rm in} = 5$, a density of
$\rho_{\rm in} = 1$, and a pressure of $p_{\rm in} = 1$. At
the coordinate origin $(x,y,z)=(0,0,0)$, a spherical source
with a radius $r = 0.1$ is established to emit gas at a rate
of $\dot{m} = 13$ units of mass per unit time with a radial
velocity of $v_r = 10$.  The outcomes of these tests are
depicted in Figure~\ref{fig:test_bow_shock}, where three
distinct cases are compared:
\begin{enumerate}
\item Uniform grid with $\Delta x = 1/32$ with full double
  precise methods;
\item Refined grid with equivalent resolution
  $\Delta x = 1/128$ within the
  $(x,y,z) \in [-1,3]\otimes[-1,1]\otimes[-1,1]$ region,
  using mixed precision methods
  (\S\ref{sec:hydro-flux-mixed});
\item Same as 2, but using full double precision methods.
\end{enumerate}
The comparisons are made at two temporal snapshots, $t = 1$
and $t = 3$. At $t = 1$, the semi-quantitative similarities
among the three cases are apparent. The first case lacks
instabilities due to higher damping from lower resolution,
while the latter two cases, by incorporating mesh
refinement, display clear signs of growing instabilities. By
$t=3$, the divergence between the latter two cases becomes
more apparent, as evidenced in the lower-right panel
illustrating the net acceleration. The reflection symmetry
across the plane $y=0$ is broken in both refined
cases, even with the utilization of full double
precision, despite commencing from identical initial
conditions that strictly maintain reflection symmetry. This
asymmetry, as discussed in \S\ref{sec:tests-lw-implode},
originates from the intricacies of floating-point
calculations on heterogeneous devices. With hydrodynamic
instability modes unattenuated at higher resolutions, the
disruption of symmetry and the ensuing chaotic turbulent
evolution are anticipated phenomena.

\begin{figure}
  \includegraphics[width=3.2in, keepaspectratio]
  {\figdir/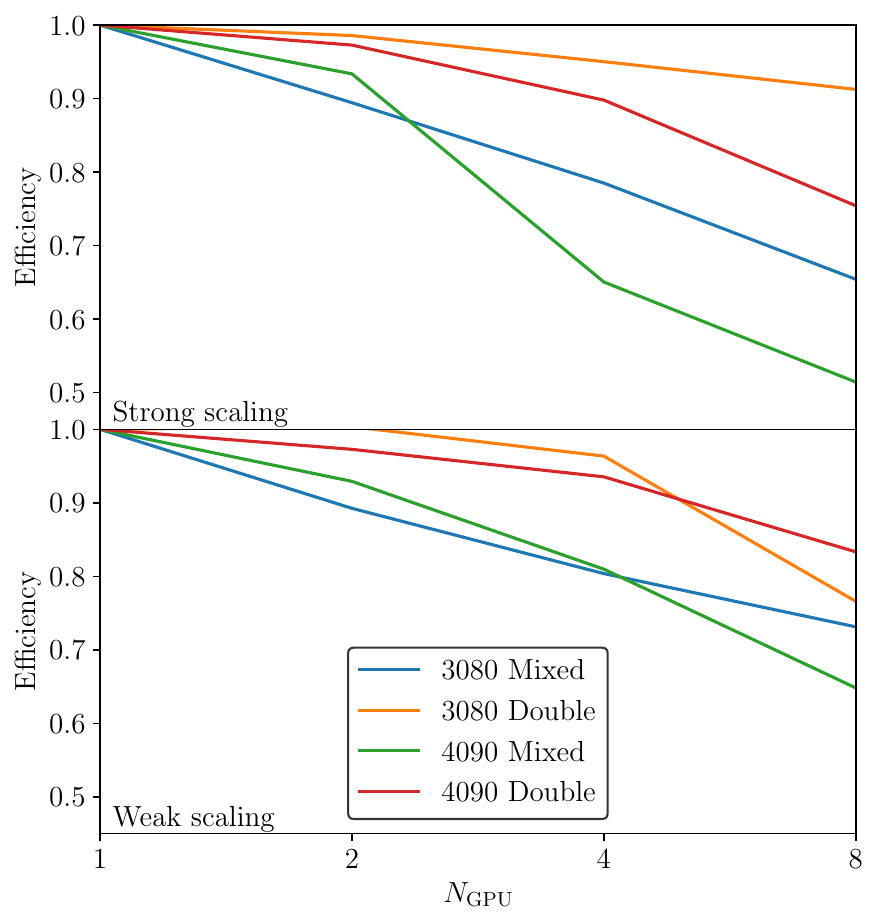}
  \caption{Scaling performance on multiple GPUs on different
    types of devices (RTX 3080 and RTX 4090), with mixed
    precision and full double precision methods
    respectively. Tests are based on the outflow--incoming
    flow interactions (see \S\ref{sec:tests-outflow-amb} and
    Figure~\ref{fig:test_bow_shock}). The upper panel shows
    the strong scaling (with the same total computation
    load), and the lower panel is for the weak scaling (with
    the same computation load per device). }
\label{fig:bs_scaling}
\end{figure}

The lower-left and lower-center panels in
Figure~\ref{fig:test_bow_shock} exhibit the application of
Hilbert curves in scenarios with uniform and refined meshes,
respectively. It is evident that with the L-system employed
in the mesh refinement framework
(\S\ref{sec:infra-load-balance}), the Hilbert curve is
properly generated within the refined region through
recursion, and task distribution is executed accordingly.  A
few scalability tests are also carried out under this
scenario, as summarized by
Figure~\ref{fig:bs_scaling}. Similar to the data presented
in \S\ref{sec:tests-rti} and Table~\ref{table:spd-rt}, the
strong and weak scalings of \kratos{} on multiple devices is
good when the single-device computation speed is relatively
slow ($\lesssim 3\times 10^8$ cells per second per device),
and the results start to deterioate with faster computing
modes (e.g., mixed precision with RTX 4090) the issue of
communication overhead (by ``not perfectly hiding
communication behind computation'').
\response{
  It is noted that the limitations on multi-GPU scalability
  are not primarily due to the host-device
  interface. Instead, the fundamental issue lies in the
  driver-level prohibition of the GPU-GPU interface for
  consumer-level GPUs, which forces code developers to
  manage communication in special ways, including those
  detailed in this works. Although \kratos{} does support
  GPU-to-GPU direct communication when it is available, this
  work deliberately avoids focusing on such optimal cases to
  prevent misleading readers. In addition, cross-node
  communication indeed results in the deterioration of
  parallelization on larger scales, with the extent of this
  deterioration largely dependent on the hardware
  setup. However, such deterioration saturates whenever the
  number of nodes employed is greater than one.  Moreover,
  introducing more microphysical modules, which is the core
  objective of constructing the entire \kratos{} system,
  considerably alleviates the issue of being unable to
  ``hide'' communication costs. These modules
  are part of subsequent manuscripts currently being
  prepared and fall outside the scope of this
  present one.
}

\section{Summary and forthcoming works}
\label{sec:summary}

This paper describes the \kratos{} Framework, a novel
GPU-based simulation system tailored for astrophysical
simulations on heterogeneous devices. \kratos{} is designed to
harness modern GPUs, providing a flexible and efficient
platform for simulating a broad spectrum of astrophysical
phenomena.  It fundamental infrastructures relies on
abstraction layers for devices, a multiprocessing
communication model, and a mesh management system. These
buildingblocks are designed for heterogeneous computing
environments (especially GPUs) consistently, preparing for
the supports for various physical modules, including
hydrodynamics (which is elaborated in this paper), and
thermochemistry, magnetohydrodynamics, self-gravity, and
radiative processes in forthcoming papers.

\kratos{} is constructed for the CUDA, HIP and similar GPU
programming models, employing abstraction layers to insulate
algorithms from programming model disparities, thereby
optimizing performance and ensuring compatibility across
different GPUs without modifying the actual implementations
of algorithms. The basic mesh structures of \kratos{}
utilizes $2^d$-trees for structured mesh management, with
mesh refinement implemented recursively. Job dispatching and
load balancing, which can involve user-defined schemes, use
Hilbert curves based on L-systems by default. Mesh
structures are designed to be separated from physical
modules, so that involving multiple physical modules can be
naturally accomplished.  The multiprocessing communication
model are implemented for device-to-device communications
that allows the GPU stream to be involved, thus optimized
for the computation on systems equipped with consumer-level
GPUs.

The hydrodynamics module of \kratos{} is constructed within
the Godunov framework, and the default implementation of
sub-modules are slope-limited Piecewise Linear Method (PLM)
reconstruction, HLLC Riemann solver, and Heun time
integrator. Optimizations for heterogeneous devices include
variable conversion, boundary conditions, flux calculation,
and time integration. The module employs mixed-precision
methods to maintain conservation laws with machine-level
accuracy while approximating fluxes.  Verification and
performance tests include Sod shock tubes, double Mach
reflection test, Liska-Wendroff implosion, Kelvin-Helmhotz
instabilities, Rayleigh-Taylor instabilities, and colliding
stellar outflows and ambients. These tests validate
algorithms, robustness, and performance of \kratos{} with results
aligning with analytic solutions and demonstrating the
effectiveness of the hydrodynamics module under various
conditions.

There are, admittedly, still prospective caveats and issues
to be solved in the current \kratos{} implementation, which
are also postponed to future works. For example, the
prohibition of direct data exchanges between consumer-level
GPUs significantly worsen the scalability of \kratos{},
especially when the computing speed on single device is too
fast to cover the communication costs. However, this issue
is naturally solved by using computing-specific GPUs that
allows faster direct communications. The inclusion of
computational heavy multiphysics modules (such as real-time
thermochemistry) will also dwarf the communication
overhead. With various modules for multiple physical
mechanisms being implemented, such as magnetohydrodynamics
solver, realtime non-equilibrium thermochemistry solver,
multigrid Poisson equation solver, particle-based solver (to
be elaborated by forthcoming papers), \kratos{} is expected to
evolve into a comprehensive system aiming at astrophysical
simulations and beyond.

\bigskip

Code availability: As \kratos{} is still being developed
actively, \response{the author will only provide the code
  upon requests and collaborations at this moment. While
  several important modules that are alerady mature have
  already been adopted and made public along with
  \citet{2025ApJS..276...40W}, a more complete version of
  \kratos{} will be available publicly after further and
  deeper debugs are accomplished.
}

\bigskip

\noindent L. Wang acknowledges the support in computing
resources provided by the Kavli Institute for Astronomy and
Astrophysics in Peking University. The author thanks the
colleagues for helpful discussions (alphabetical order):
Xue-Ning Bai, Renyue Cen, Zhuo Chen, Can Cui, Ruobin Dong,
Jeremy Goodman, Luis C. Ho, Xiao Hu, Kohei Inayoshi,
Mordecai Mac-Low, Ming Lv, Kengo Tomida, Zhenyu Wang,
Haifeng Yang, and Yao Zhou, for helpful discussions and
useful suggestions.

\bibliographystyle{aasjournal}
\bibliography{method}


\end{document}